\documentclass[letterpaper,twocolumn,10pt]{article}
\usepackage{usenix2019_v3}

\usepackage{amsmath}
\usepackage{amsfonts}
\usepackage{graphicx}
\usepackage{subcaption}
\usepackage{enumitem}

\usepackage{soul}
\usepackage{textcomp}
\usepackage{multirow}
\usepackage{tablefootnote}

\newcommand{\sys}{\textit{TCPlp}}
\newcommand{\cwnd}{\texttt{cwnd}}
\newcommand{\ssthresh}{\texttt{ssthresh}}
\newcommand{\mss}{\text{MSS}}
\newcommand{\rtt}{\text{RTT}}
\newcommand{\eto}{\text{ETO}}
\newcommand{\pwin}{p_{\text{win}}}
\newcommand{\trec}{t_{\text{rec}}}

\newcommand{\secref}[1]{\S{}\ref{#1}}
\newcommand{\appref}[1]{Appendix \ref{#1}}
\newcommand{\parhead}[1]{\noindent\textbf{#1.}}

\usepackage{titlesec}
\titlespacing*\section{0pt}{3pt plus 2pt minus 2pt}{1pt plus 1pt minus 1pt}
\titlespacing*\subsection{0pt}{2pt plus 2pt minus 1pt}{1pt plus 1pt minus 1pt}
\titlespacing*\subsubsection{0pt}{2pt plus 2pt minus 1pt}{1pt plus 1pt minus 1pt}

\usepackage{titling}
\begin{document}

\date{}

\title{\Large \bf Performant TCP for Low-Power Wireless Networks}

\author{
{\rm Sam Kumar, Michael P Andersen, Hyung-Sin Kim, and David E. Culler}\\
\textit{University of California, Berkeley} 
} 

\maketitle

\begin{abstract}
Low-power and lossy networks (LLNs) enable diverse applications integrating many resource-constrained embedded devices, often requiring interconnectivity with existing TCP/IP networks as part of the Internet of Things. But TCP has received little attention in LLNs due to concerns about its overhead and performance, leading to LLN-specific protocols that require specialized gateways for interoperability. We present a systematic study of a well-designed TCP stack in IEEE 802.15.4-based LLNs, based on the TCP protocol logic in FreeBSD. Through careful implementation and extensive experiments, we show that modern low-power sensor platforms are capable of running full-scale TCP and that TCP, counter to common belief, performs well despite the lossy nature of LLNs. By carefully studying the interaction between the transport and link layers, we identify subtle but important modifications to both, achieving TCP goodput within 25\% of an upper bound (5--40x higher than prior results) and low-power operation commensurate to CoAP in a practical LLN application scenario. This suggests that a TCP-based transport layer, seamlessly interoperable with existing TCP/IP networks, is viable and performant in LLNs.

\end{abstract}

\section{Introduction}\label{sec:intro}

Research on wireless networks of low-power, resource constrained, embedded devices---low-power and lossy networks (LLNs) in IETF terms~\cite{vasseur2014rfc}---blossomed in the late 1990s. To obtain freedom to tackle the unique challenges of LLNs, researchers initially departed from the established conventions of the Internet architecture~\cite{estrin1999next, hill2000system}. As the field matured, however, researchers found ways to address these challenges \emph{within} the Internet architecture~\cite{hui2008ip}.
Since then, it has become commonplace to use IPv6 in LLNs via the 6LoWPAN~\cite{montenegro2008lowpan} adaptation layer.
IPv6-based routing protocols, like RPL~\cite{winter2012rpl}, and application-layer transports over UDP, like CoAP~\cite{castellani2011web}, have become standards in LLNs.
Most wireless sensor network (WSN) operating systems, such as TinyOS~\cite{levis2005tinyos}, RIOT~\cite{baccelli2018riot}, and Contiki~\cite{dunkels2004contiki}, ship with IP implementations enabled and configured. Major industry vendors offer branded and supported 6LoWPAN stacks (e.g., TI SimpleLink, Atmel SmartConnect).
A consortium, Thread~\cite{thread}, has formed around 6LoWPAN-based interoperability.

Despite these developments, transport in LLNs has remained ad-hoc and TCP has received little serious consideration. Many embedded IP stacks (e.g., OpenThread~\cite{openthread}) do not even support TCP, and those that do implement only a subset of its features (Appendix \ref{sec:features}). The conventional wisdom is that IP holds merit, but \textit{TCP is ill-suited to LLNs}. This view is represented by concerns about TCP, such as:

\begin{itemize}[leftmargin=*, noitemsep, topsep=0mm]
    \item ``TCP is not light weight ... and may not be suitable for implementation in low-cost sensor nodes with limited processing, memory and energy resources.''~\cite{pang2007reliable} (Similar argument in \cite{dunkels2003full}, \cite{iyer2005stcp}.)
    \item That ``TCP is a connection-oriented protocol'' is a poor match for WSNs, ``where actual data might be only in the order of a few bytes.''~\cite{rahman2008wireless} (Similar argument in \cite{pang2007reliable}.)
    \item ``TCP uses a single packet drop to infer that the network is congested.'' This ``can result in extremely poor transport performance because wireless links tend to exhibit relatively high packet loss rates.''~\cite{paek2007rcrt} (Similar argument in \cite{dunkels2003making}, \cite{dunkels2004connecting}, \cite{iyer2005stcp}.)
\end{itemize}
Such viewpoints have led to a plethora of WSN-specialized protocols and systems~\cite{pang2007reliable, rathnayaka2013wireless, wang2005issues} for reliable data transport, such as
PSFQ~\cite{wan2002psfq},
STCP~\cite{iyer2005stcp},
RCRT~\cite{paek2007rcrt},
Flush~\cite{kim2007flush},
RMST~\cite{stann2003rmst},
Wisden~\cite{xu2004wisden},
CRRT~\cite{alam2009crrt}, and
CoAP~\cite{bormann2012coap},
and for unreliable data transport, like
CODA~\cite{wan2003coda},
ESRT~\cite{sankarasubramaniam2003esrt},
Fusion~\cite{hull2004mitigating},
CentRoute~\cite{stathopoulos2005mote},
Surge~\cite{levis2003tossim}, and
RBC~\cite{zhang2005reliable}.

As LLNs become part of the emerging Internet of Things (IoT), it behooves us to re-examine the transport question, with attention to how the landscape has shifted:
(1) As part of IoT, LLNs must be interoperable with traditional TCP/IP networks; to this end, using TCP in LLNs simplifies IoT gateway design.
(2) Popular IoT application protocols, like MQTT~\cite{mqtt} and ZeroMQ~\cite{zeromq}, assume that TCP is used at the transport layer.
(3) Some IoT application scenarios demand high link utilization and reliability on low-bandwidth lossy links.
Embedded hardware has also evolved substantially, prompting us to revisit TCP's overhead.
In this context, \textbf{this paper seeks to determine: Do the ``common wisdom'' concerns about TCP hold in a modern IEEE 802.15.4-based
LLN? Is TCP (still) unsuitable for use in LLNs?}

To answer this question, we leverage the fully-featured TCP implementation in the FreeBSD Operating System (rather than a limited locally-developed implementation) and refactor it to work with the Berkeley Low-Power IP Stack (BLIP), Generic Network Stack (GNRC), and OpenThread network stack, on two modern LLN platforms (\secref{sec:implementation}). Na\"ively running TCP in an LLN indeed results in poor performance. However, upon close examination, we discover that this is not caused by the expected reasons, such as those listed above. The \emph{actual} reasons for poor TCP performance include (1) small link-layer frames that increase TCP header overhead, (2) hidden terminal effects over multiple wireless hops, and (3) poor interaction between TCP and a duty-cycled link.
Through a systematic study of TCP in LLNs, we develop techniques to resolve these issues (Table \ref{tab:techniques}), uncover why the generally assumed problems do not apply to TCP in LLNs, and show that TCP perfoms well in LLNs once these issues are resolved:

\begin{table}[t]
    \centering
    \setlength\tabcolsep{1.5pt}
    \begin{tabular}{|c|c|c|}\hline
        Challenge & Technique & Observed Improvement \\\hline\hline
        Resource & Zero-Copy Send & Send Buffer: 50\% less mem.\\\cline{2-3}
        Constraints & In-Place Reass. & Recv Buffer: 38\% less mem.\\\hline\hline
        Link-Layer & Large MSS & TCP Goodput: 4--5x higher\\\cline{2-3}
        Properties & Link Retry Delay & TCP Seg. Loss: $6\% \rightarrow 1\%$\\\hline\hline
        Energy & Adaptive DC & HTTP Latency: $\approx 2$x lower\\\cline{2-3}
        Constraints & L2 Queue Mgmt. & TCP Radio DC: $3\% \rightarrow 2\%$\\\hline
    \end{tabular}
    \caption{Impact of techniques to run full-scale TCP in LLNs}
    \label{tab:techniques}
\end{table}

We find that \textbf{full-scale TCP fits well within the CPU and memory constraints of modern LLN platforms} (\secref{sec:implementation}, \secref{sec:interactions}). Owing to the low bandwidth of a low-power wireless link, a small window size ($\approx 2$ KiB) is sufficient to fill the bandwidth-delay product and achieve good TCP performance. This translates into small send/receive buffers that fit comfortably within the memory of modern WSN hardware.
Furthermore, we propose using an atypical Maximum Segment Size (MSS) to manage header overhead and packet fragmentation.
As a result, \textbf{full-scale TCP operates well in LLNs, with 5--40 times higher throughput than existing (relatively simplistic) embedded TCP stacks} (\secref{sec:interactions}).

Hidden terminals are a serious problem when running TCP over multiple wireless hops. We propose adding a delay $d$ between link-layer retransmissions, and demonstrate that it effectively reduces hidden-terminal-induced packet loss for TCP.
We find that, because a small window size is sufficient for good performance in LLNs, \textbf{TCP is quite resilient to spurious packet losses, as the congestion window can recover to a full window quickly after loss} (\secref{sec:multihop}).

To run TCP in a low-power context, we \emph{adaptively} duty-cycle the radio to avoid poor interactions with TCP's self-clocking behavior. We also propose careful link-layer queue management to make TCP more robust to interference. We demonstrate that \textbf{TCP can operate at low power, comparable to alternatives tailored specifically for WSNs}, and that \textbf{TCP brings value for real IoT sensor applications} (\secref{sec:application}).

We conclude that TCP is entirely capable of running on IEEE 802.15.4 networks and low-cost embedded devices in LLN application scenarios (\secref{sec:conclusion}). Since our improvements to TCP and the link layer maintain seamless interoperability with other TCP/IP networks, we believe that a TCP-based transport architecture for LLNs could yield considerable benefit.

In summary, this paper's contributions are:
\begin{enumerate}[leftmargin=*,noitemsep,topsep=0ex]
    \item We implement a full-scale TCP stack for low-power embedded devices and reduce its resource usage.
    \item We identify the actual issues causing poor TCP performance and develop techniques to address them.
    \item We explain why the expected insurmountable reasons for poor TCP performance actually do not apply.
    \item We demonstrate that, once these issues are resolved, TCP performs comparably to LoWPAN-specialized protocols.
\end{enumerate}
Table \ref{tab:techniques} lists our techniques to run TCP in an LLN.
Although prior LLN work has already used various forms of link-layer delays~\cite{woo2001transmission} and adaptive duty-cycling~\cite{ye2004medium}, our work shows, where applicable, how to adapt these techniques to work well with TCP, and demonstrates that they can address the challenges of LLNs within a \emph{TCP-based} transport architecture.

\section{Background and Related Work}\label{sec:background}

Since the introduction of TCP, a vast literature has emerged, focusing on improving it as the Internet evolved. Some representative areas include congestion control~\cite{jacobson1988congestion, fall1996simulation, grieco2004performance, afanasyev2010host},
performance on wireless links~\cite{balakrishnan1995improving, allman1999enhancing},
performance in high-bandwidth environments~\cite{borman2014tcp, floyd2003highspeed, jin2004fast, ha2008cubic, alizadeh2010data},
mobility~\cite{snoeren2000end},
and multipath operation~\cite{raiciu2012mptcp}.
Below, we discuss TCP in the context of LLNs and embedded devices.

\subsection{Low-Power and Lossy Networks (LLNs)}

Although the term \emph{LLN} can be applied to a variety of technologies, including LoRa and Bluetooth Low Energy, we restrict our attention in this paper to \textbf{embedded networks using IEEE 802.15.4}. Such networks are called LoWPANs~\cite{kushalnagar2007rfc}---Low-Power Wireless Personal Area Networks---in contrast to WANs, LANs (802.3), and WLANs (802.11).
Outside of LoWPANs, TCP has been successfully adapted to a variety of networks, including
serial~\cite{jacobson1990compressing},
Wi-Fi~\cite{balakrishnan1995improving},
cellular~\cite{balakrishnan1997effects, ludwig2003tcp},
and
satellite~\cite{balakrishnan1997effects, allman1999enhancing} links.
While an 802.15.4 radio can in principle be added as a NIC to any device, we consider only \emph{embedded} devices where it is the primary means of communication, running operating systems like TinyOS~\cite{hill2000system}, RIOT~\cite{baccelli2018riot}, Contiki~\cite{dunkels2004contiki}, or FreeRTOS. These devices are currently built around microcontrollers with Cortex-M CPUs, which lack MMUs.
Below, we explain how LoWPANs are different from other networks where TCP has been successfully adapted.

\smallskip
\parhead{Resource Constraints}
When TCP was initially adopted by ARPANET in the early 1980s, contemporary Internet citizens---typically minicomputers and high-end workstations, but not yet personal computers---usually had at least 1 MiB of RAM. 1 MiB is tiny by today's standards, yet the LLN-class devices we consider in this work have \emph{1-2 orders of magnitude less RAM than even the earliest computers connected with TCP/IP}.
Due to energy constraints, particularly SRAM leakage, RAM size in low-power MCUs does not follow Moore's Law.
For example, comparing Hamilton~\cite{kim2018system}, which we use in this work, to TelosB~\cite{polastre2005telos}, an LLN platform from 2004, shows only a 3.2x increase in RAM size over 16 years.
This has caused LLN-class embedded devices to have a different balance of resources than conventional systems, a trend that is likely to continue well into the future.
For example, whereas conventional computers have historically had roughly 1 MiB of RAM for every MIPS of CPU, as captured by the 3M rule, Hamilton has $\approx 50$ DMIPS of CPU but only 32 KiB of RAM.

\smallskip
\parhead{Link-Layer Properties}
IEEE 802.15.4 is a low-bandwidth, wireless link with an MTU of only 104 bytes. The research community has explored using TCP with links that are \emph{separately} low-bandwidth, wireless~\cite{balakrishnan1995improving}, or low-MTU~\cite{jacobson1990compressing}, but addressing these issues \emph{together} raises new challenges.
For example, RTS-CTS, used in WLANs to avoid hidden terminals, has high overhead in LoWPANs~\cite{woo2001transmission, hull2004mitigating} due to the small MTU---control frames are comparable in size to data frames.
Thus, LoWPAN researchers have moved away from RTS-CTS, instead carefully designing application traffic patterns to avoid hidden terminals~\cite{kim2007flush, polastre2004versatile, hull2004mitigating}.
Unlike Wi-Fi/LTE, LoWPANs do not use physical-layer techniques like adaptive modulation/coding or multi-antenna beamforming. Thus, they are directly impacted by link quality degradation due to varying environmental conditions~\cite{szewczyk2004lessons, polastre2004versatile}. Additionally, IEEE 802.15.4 coexists with Wi-Fi in the 2.4 GHz frequency band, making Wi-Fi interference particularly relevant in indoor settings~\cite{liang2010surviving}. As LoWPANs are \emph{embedded} networks, there is no human in the loop to react to and repair bad link quality.

\smallskip
\parhead{Energy Constraints}
Embedded nodes---the ``hosts'' of an LLN---are subject to strict power constraints.
Low-power radios consume almost as much energy listening for a packet as they do when actually sending or receiving~\cite{kim2018system, at86rf233}. Therefore, it is customary to \emph{duty-cycle} the radio, keeping it in a low-power sleep state, in which it cannot send or receive data, most of the time~\cite{ye2002energy, polastre2004versatile, hui2008ip}. The radio is only \emph{occasionally} turned on to send/receive packets or determine if reception is likely.
This requires \emph{Media Management Control (MMC)} protocols~\cite{ye2002energy, polastre2004versatile, hui2008ip} at the link layer to ensure that frames destined for a node are delivered to it only when its radio is on and listening.
Similarly, the CPU also consumes a significant amount of energy~\cite{kim2018system}, and must be kept idle most of the time.

Over the past 20 years, LLN researchers have addressed these challenges, but only in the context of special-purpose networks highly tailored to the particular application task at hand.
The remaining open question is how to do so with a general-purpose reliable transport protocol like TCP.

\subsection{TCP/IP for Embedded LLN-Class Devices}

In the late 1990s and early 2000s, developers attempted to bring TCP/IP to embedded and resource-constrained systems to connect them to the Internet, usually over serial or Ethernet. Such systems~\cite{borriello2000embedded, ju2000efficient} were often designed with a specific application---often, a web server---in mind.
These TCP/IP stacks were tailored to the specific applications at hand and were not suitable for general use.
uIP (``micro IP'')~\cite{dunkels2003full}, introduced in 2002, was a standalone \emph{general} TCP/IP stack optimized for 8-bit microcontrollers and serial or Ethernet links. To minimize resource consumption to run on such platforms, uIP omits standard features of TCP; for example, it allows only a single outstanding (unACKed) TCP segment per connection, rather than a sliding window of in-flight data.

Since the introduction of uIP, embedded networks have changed substantially. With \emph{wireless} sensor networks and IEEE 802.15.4, various low-power networking protocols have been developed to overcome lossy links with strict energy and resource constraints, from
S-MAC~\cite{ye2002energy}, B-MAC~\cite{polastre2004versatile}, X-MAC~\cite{buettner2006x}, and A-MAC~\cite{dutta2010design}, to
Trickle~\cite{levis2004trickle} and CTP~\cite{gnawali2009collection}.
Researchers have viewed TCP as unsuitable, however, questioning end-to-end recovery, loss-triggered congestion control, and bi-directional data flow in  LLNs~\cite{dunkels2004connecting}.
Furthermore, WSNs of this era typically did not even use IP; instead, each WSN was designed specifically to support a particular application~\cite{mainwaring2002wireless, xu2004wisden, kim2007health}. Those that require global connectivity rely on application-specific ``base stations'' or ``gateways'' connected to a TCP/IP network, treating the LLN like a peripheral interconnect (e.g., USB, bluetooth) rather than a network in its own right. This is because the prevailing sentiment at the time was that LLNs are too different from other types of networks and have to operate in too extreme conditions for the layered Internet architecture to be appropriate~\cite{estrin1999next}.

In 2007, the 6LoWPAN adaptation layer~\cite{montenegro2008lowpan} was introduced, enabling IPv6 over IEEE 802.15.4. IPv6 has since been adopted in LLNs, bringing forth IoT~\cite{hui2008ip}.
uIP has been ported to LLNs~\cite{durvy2008making}, and
IPv6 routing protocols, like RPL~\cite{winter2012rpl}, and UDP-based application-layer transports, like CoAP~\cite{castellani2011web}, have emerged in LLNs.
Representative operating systems, like TinyOS and Contiki, implement UDP/RPL/IPv6/6LoWPAN network stacks with IEEE 802.15.4-compatible MMC protocols for 16-bit platforms like TelosB~\cite{polastre2005telos}.

TCP, however, is not widely adopted in LLNs.
The few LLN studies that use TCP~\cite{duquennoy2011lossy, hewage2015enabling, hui2008ip, im2015tcp, kim2015measurement, zheng2011tcp, gomez2018tcp}
generally use a simplified TCP stack (\appref{sec:features}), such as uIP.

In summary, despite the acceptance of IPv6, LLNs remain highly tailored at the transport layer to the application at hand. They typically use application-specific protocols on top of UDP; of such protocols, CoAP~\cite{bormann2012coap} has the widest adoption. In this context, this paper explores whether adopting TCP---and more broadly, the ecosystem of IP-based protocols, rather than IP alone---might bring value to LLNs moving forward.

\section{Motivation: The Case for TCP in LLNs}\label{sec:tcp}

As explained in \secref{sec:background}, LLN design has historically been highly tailored to the specific application task at hand, for maximum efficiency. For example, PSFQ broadcasts data from a single source node to all others, RMST supports ``directed diffusion''~\cite{intanagonwiwat2000directed}, and CoAP is tied to REST semantics. But embedded networks are not just isolated devices (e.g., peripheral interconnects like USB or bluetooth)---they are now true Internet citizens, and should be designed as such.

In particular, the recent megatrend of IoT requires LLNs to have a greater degree of \emph{interoperability} with regular TCP/IP networks. Yet, LLN-specific protocols lack a clear separation between the transport and application layers, requiring \emph{application-layer gateways} to communicate with TCP/IP-based services. This has encouraged IoT applications to develop as vertically-integrated silos, where devices cooperate only within an individual application or a particular manufacturer's ecosystem, with little to no interoperability \emph{between} applications or with the general TCP/IP-based Internet.
This phenomenon, sometimes called the ``CompuServe of Things,'' is a serious obstacle to the IoT vision~\cite{zachariah2015internet, levy2016beetle, furst2016leveraging, mcewen2013risking, windley2014compuserve}. In contrast, other networks are seamlessly interoperable with the rest of the Internet. Accessing a new web application from a laptop does not require any new functionality at the Wi-Fi access point, but running a new application in a gateway-based LLN \emph{does} require additional application-specific functionality to be installed at the gateway.

In this context, TCP-enabled LLN devices would be first-class citizens of the Internet, natively interoperable with the rest of the Internet via TCP/IP. They could use IoT protocols that assume a TCP-based transport layer (e.g., MQTT~\cite{mqtt}) and security tools for TCP/IP networks (e.g., stateful firewalls), without an application-layer gateway.
In addition, while traditional LLN applications like environment monitoring can be supported by unreliable UDP, certain applications do require high throughput and reliable delivery (e.g., anemometry (\appref{app:anemometer}), vibration monitoring~\cite{jung2017vibration}). TCP, \textit{if it performs well in LLNs}, could benefit these applications.

Adopting TCP in LLNs may also open an interesting research agenda for IoT. TCP is the default transport protocol outside of LLNs, and history has shown that, to justify other transport protocols, application characteristics must offer substantial opportunity for optimization (e.g., \cite{winstein2012mosh, winstein2013stochastic, fouladi2018salsify}). If TCP becomes a viable option in LLNs, it would raise the bar for application-specific LLN protocols, resulting in some potentially interesting alternatives.

Although adopting TCP in LLNs could yield significant benefit and an interesting agenda, its feasibility and performance remain in question. This motivates our study.

\section{Empirical Methodology}

This section presents our methodology, carefully chosen to ground our study of full-scale TCP in LLNs.

\subsection{Network Stack}\label{sec:thread}
\parhead{Transport layer}
That only a few full-scale TCP stacks exist, with a body of literature covering decades of refining, demonstrates that developing a feature-complete implementation of TCP is complex and error-prone~\cite{paxson1999known}. Using a well-tested TCP implementation would ensure that results from our measurement study are due to the TCP \emph{protocol}, not an artifact of the TCP \emph{implementation} we used. Thus, we leverage the TCP implementation in FreeBSD 10.3~\cite{freebsd} to ground our study. We ported it to run in embedded operating systems and resource-constrained embedded devices (\secref{s:hardware}).

To verify the effectiveness of full-scale TCP in LLNs, we compare with CoAP~\cite{shelby2014constrained}, CoCoA~\cite{betzler2016coap}, and unreliable UDP.
CoAP is a standard LLN protocol that provides reliability on top of UDP.
It is the most promising LLN alternative to TCP, gaining momentum in both academia~\cite{colitti2011evaluation, villaverde2012constrained, kovatsch2014californium, santos2015personal, betzler2016coap, seitz2017enabling} and
industry~\cite{eclipse2014coap, oma2016constrained}, with adoption by Cisco~\cite{cisco2013beyond, cisco2018software}, Nest/Google~\cite{openthreadwebsite}, and Arm~\cite{armjava, armdevicemanagementconnect}.
CoCoA~\cite{betzler2016coap} is a recent proposal that augments CoAP with RTT estimation.

It is attractive to compare TCP to a variety of commercial systems, as has been done by a number of studies in LTE/WLANs~\cite{winstein2013stochastic, fouladi2018salsify}. Unfortunately, multihop LLNs have not yet reached the level of maturity to support a variety of commercial offerings; only CoAP has an appreciable level of commercial adoption.
Other protocols are research proposals that often
(1) are implemented for now-outdated operating systems and hardware or exist only in simulation~\cite{iyer2005stcp, kim2007flush, alam2009crrt},
(2) target a very specific application paradigm~\cite{wan2002psfq, stann2003rmst, xu2004wisden}, and/or
(3) do not use IP~\cite{wan2002psfq, iyer2005stcp, kim2007flush, paek2007rcrt}.
We choose CoAP and CoCoA because they are not subject to these constraints.

\smallskip
\parhead{Layers 1 to 3}
Because it is burdensome to place a border router with LAN connectivity within wireless range of every low-power host (e.g., sensor node), it is common to transmit data (e.g., readings) over \emph{multiple} wireless LLN hops.
Although each sensor must be battery-powered, it is reasonable to have a wall-powered LLN router node within transmission range of it.\footnote{The assumption of powered ``core routers'' is reasonable for most IoT use cases, which are typically indoors. Recent IoT protocols, such as Thread~\cite{thread} and BLEmesh~\cite{blemesh}, take advantage of powered core routers.}
This motivates Thread\footnote{Thread has a large amount of industry support with a consortium already consisting of over 100 members~\cite{threadmembers}, and is used in real IoT products sold by Nest/Google~\cite{threadproducts}.
Given this trend, using Thread makes our work timely.}~\cite{thread, kim2019thread}, a recently developed protocol standard that constructs a multihop LLN over IEEE 802.15.4 links with \emph{wall-powered, always-on} router nodes and \emph{battery-powered, duty-cycled}
leaf nodes.
We use OpenThread~\cite{openthread}, an open-source implementation of Thread.

Thread decouples routing from energy efficiency, providing a full-mesh topology among routers, frequent route updates, and asymmetric bidirectional routing for reliability.
Each \emph{leaf node} duty cycles its radio, and simply chooses a core router with good link quality, called its \emph{parent}, as its next hop to all other nodes.
The duty cycling uses \emph{listen-after-send}~\cite{schmid2010disentangling}. A leaf node's parent stores downstream packets destined for that leaf node, until the leaf node sends it a \emph{data request} message. A leaf node, therefore, can keep its radio powered off most of the time; infrequently, it sends a data request message to its parent, and turns on its radio for a short interval afterward to listen for downstream packets queued at its parent. Leaf nodes may send upstream traffic at any time. Each node uses CSMA-CA for medium access.

\begin{table}[t]
    \centering
    \setlength\tabcolsep{3.5pt}
    \begin{tabular}{| l || c | c | c | c |}\hline
         & TelosB & Hamilton & Firestorm & Raspberry Pi \\\hline\hline
        CPU & MSP430 & Cortex-M0+ & Cortex-M4 & Cortex-A53\\\hline
        RAM & 10 KiB & 32 KiB & 64 KiB & 256 MB\\\hline
        ROM & 48 KiB & 256 KiB & 512 KiB & SD Card\\\hline
    \end{tabular}
    \caption{Comparison of the platforms we used (Hamilton and Firestorm) to TelosB and Raspberry Pi}
    \label{table:platform_comparison}
\end{table}

\subsection{Embedded Hardware}\label{s:hardware}
We use two embedded hardware platforms: Hamilton~\cite{kim2018system} and Firestorm~\cite{andersen2016system}. Hamilton uses a SAMR21 SoC with a 48 MHz Cortex-M0+, 256 KiB of ROM, and 32 KiB of RAM. Firestorm uses a SAM4L 48 MHz Cortex-M4 with 512 KiB of ROM and 64 KiB of RAM.
While these platforms are more powerful than the TelosB~\cite{polastre2005telos}, an older LLN platform widely used in past studies, they are heavily resource-constrained compared to a Raspberry Pi (Table \ref{table:platform_comparison}).
Both platforms use the AT86RF233 radio, which supports IEEE 802.15.4. We use its standard data rate, 250 kb/s.
We use Hamilton/OpenThread in our experiments; for comparison, we provide some results from Firestorm and other network stacks in \appref{sec:netstack}.

\smallskip
\parhead{Handling automatic radio features}
The AT86RF233 radio has built-in hardware support for link-layer retransmissions and CSMA-CA. However, it automatically enters low-power mode during CSMA backoff, during which it does not listen for incoming frames~\cite{at86rf233}. This behavior, which we call \emph{deaf listening}, interacts poorly with TCP when radios are always on, because TCP requires bidirectional flow of packets---data in one direction and ACKs in the other. This may initially seem concerning, as deaf listening is an important power-saving feature. Fortunately, this problem disappears when using OpenThread's listen-after-send duty-cycling protocol, as leaf nodes never transmit data when listening for downstream packets.
For experiments with always-on radios, we do not use the radio's capability for hardware CSMA and link retries; instead, we perform these operations in software.

\begin{figure}[t]
    \centering
    \includegraphics[width=\linewidth]{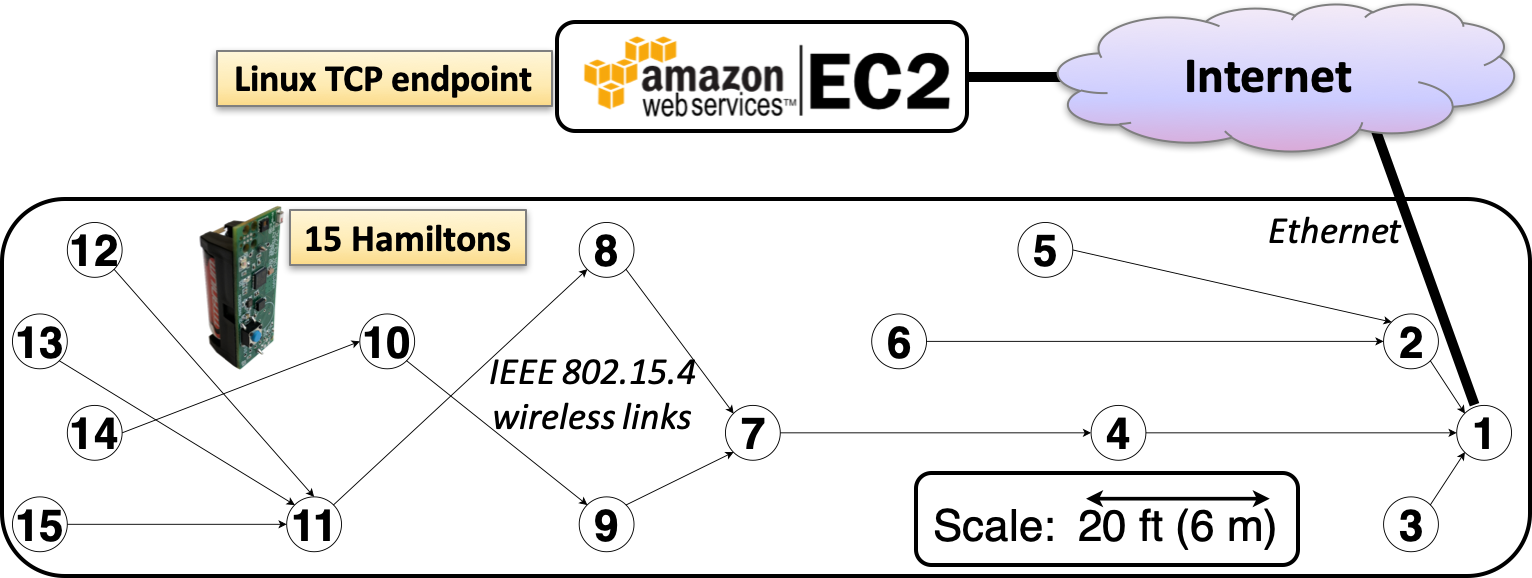}
    \caption{Snapshot of uplink routes in OpenThread topology at transmission power of -8 dBm (5 hops). Node 1 is the border router with Internet connectivity.}
    \label{fig:topology}
\end{figure}

\smallskip
\parhead{Multihop Testbed}\label{s:testbed}
We construct an indoor LLN testbed, depicted in Figure \ref{fig:topology}, with 15 Hamiltons where node 1 is configured as the border router. OpenThread forms a 3-to-5-hop topology at transmission power of -8 dBm. Embedded TCP endpoints (Hamiltons) communicate with a Linux TCP endpoint (server on Amazon EC2) via the border router.
During working hours, interference is present in the channel, due to people in the space using Wi-Fi and Bluetooth devices in the 2.4 GHz frequency band. At night, when there are few/no people in the space, there is much less interference.

\section{Implementation of \sys{}}\label{sec:implementation}

We seek to answer the following two questions: (1) Does full-scale TCP fit within the limited memory of modern LLN platforms? (2) How can we integrate a TCP implementation from a traditional OS into an embedded OS?
To this end, we develop a TCP stack for LLNs based on the TCP implementation in FreeBSD 10.3, called \sys{}~\cite{kumar2018bringing}, on multiple embedded operating systems, RIOT OS~\cite{baccelli2018riot} and TinyOS~\cite{levis2005tinyos}.
We use \sys{} in our measurement study in future sections.

Although we carefully preserved the protocol logic in the FreeBSD TCP implementation, achieving correct and performant operation on sensor platforms was a nontrivial effort.
We had to modify the FreeBSD implementation according to the concurrency model of each embedded network stack and the timer abstractions provided by each embedded operating system (\appref{sec:netstack}).
Our other modifications to FreeBSD, aimed at reducing memory footprint, are described below.

\subsection{Connection State for \sys{}}

As discussed in \appref{sec:features}, \sys{} includes features from FreeBSD that improve standard communication, like a sliding window, New Reno congestion control, zero-window probes, delayed ACKs, selective ACKs, TCP timestamps, and header prediction.
\sys{}, however, omits some features in FreeBSD's TCP/IP stack. We omit dynamic window scaling, as buffers large enough to necessitate it ($\geq 64$ KiB) would not fit in memory. We omit the urgent pointer, as it not recommended for use~\cite{gont2011implementation} and would only complicate buffering. Certain security features, such as host cache, TCP signatures, SYN cache, and SYN cookies are outside the scope of this work. We do, however, retain challenge ACKs~\cite{ramaiah2010improving}.

We use separate structures for \emph{active sockets} used to send and receive bytes, and \emph{passive sockets} used to listen for incoming connections, as passive sockets require less memory.

Table \ref{table:memory_riot} depicts the memory footprint of \sys{} on RIOT OS.
The memory required for the protocol and application state of an active TCP socket fits in a few hundred bytes, less than 1\% of the available RAM on the Cortex-M4 (Firestorm) and 2\% of that on the Cortex-M0+ (Hamilton). Although \sys{} includes heavyweight features not traditionally included in embedded TCP stacks, it fits well within available memory.

\begin{table}[t]
    \centering
    \setlength\tabcolsep{3pt}
    \begin{tabular}{| l || c | c | c |}\hline
         & Protocol & Socket Layer & \texttt{posix\_sockets} \\\hline\hline
         ROM & 19972 B & 6216 B & 5468 B \\\hline
         RAM (Active) & 364 B & 88 B & 48 B \\\hline
         RAM (Passive) & 12 B & 88 B & 48 B\\\hline
    \end{tabular}
    \caption{Memory usage of \sys{} on RIOT OS. We also include RIOT's \texttt{posix\_sockets} module, used by \sys{} to provide a Unix-like interface.}
    \label{table:memory_riot}
\end{table}

\subsection{Memory-Efficient Data Buffering}\label{sec:buffering}

Existing embedded TCP stacks, such as uIP and BLIP, allow \textit{only one TCP packet in the air}, eschewing careful implementation of send and receive buffers~\cite{kim2015measurement}. These buffers, however, are key to supporting TCP's sliding window functionality. We observe in \secref{sec:window} that \sys{} performs well with only 2-3 KiB send and receive buffers, which comfortably fit in memory even when na\"ively pre-allocated at compile time. Given that buffers dominate \sys{}'s memory usage, however, we discuss techniques to optimize their memory usage.

\subsubsection{Send Buffer: Zero-Copy}\label{s:zerocopy}

Zero-copy techniques~\cite{bershad1989lrpc, druschel1993fbufs, khalidi1995zerocopy, li2005lyranet, maeda1993protocol} were devised for situations where the time for the CPU to copy memory is a significant bottleneck. Our situation is very different; the radio, not the CPU, is the bottleneck, owing to the low bandwidth of IEEE 802.15.4. By using a zero-copy send buffer, however, we can avoid allocating memory to intermediate buffers that would otherwise be needed to copy data, thereby reducing the network stack's total memory usage.

In TinyOS, for example, the BLIP network stack supports vectored I/O; an outgoing packet passed to the IPv6 layer is specified as an \texttt{iovec}. Instead of allocating memory in the packet heap for each outgoing packet,
\sys{} simply creates \texttt{iovec}s that point to existing data in the send buffer. This decreases the required size of the packet heap.

Unfortunately, zero-copy optimizations were not possible for the OpenThread implementation, because OpenThread does not support vectored I/O for sending packets. The result is that the \sys{} implementation requires a few kilobytes of additional memory for the send buffer on this platform.

\subsubsection{Receive Buffer: In-Place Reassembly Queue}\label{s:reassembly}

Not all zero-copy optimizations are useful in the embedded setting.
In FreeBSD, received packets are passed to the TCP implementation as \texttt{mbuf}s~\cite{wright1995tcpvol2chap2}.
The receive buffer and reassembly buffer are \texttt{mbuf} chains, so data need not be copied out of \texttt{mbuf}s to add them to either buffer or recover from out-of-order delivery.
Furthermore, buffer sizes are chosen dynamically~\cite{semke1998automatic}, and are merely a \emph{limit} on their actual size.
In our memory-constrained setting, such a design is dangerous because its memory usage is nondeterministic; there is additional memory overhead, due to headers, if the data are delivered in many small packets instead of a few large ones.

\begin{figure}[t]
    \centering
    \begin{subfigure}[p]{\linewidth}
        \includegraphics[width=\linewidth]{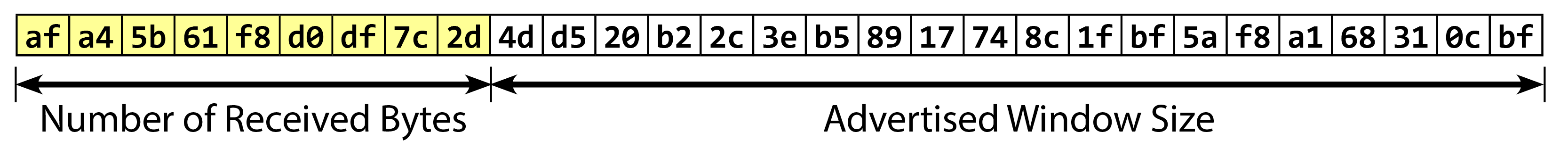}
        \caption{Na\"ive receive buffer. Note that $\text{size of advertised window} + \text{size of buffered data} = \text{size of receive buffer}$.}
        \label{fig:receive_buffer_naive}
    \end{subfigure}
    \begin{subfigure}[p]{\linewidth}
        \includegraphics[width=\linewidth]{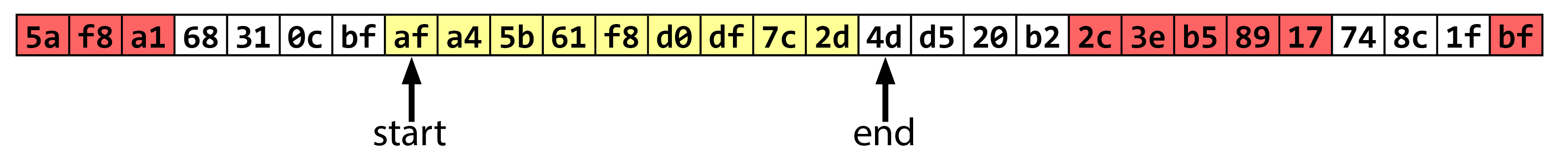}
        \caption{Receive buffer with in-place reassembly queue. In-sequence data (yellow) is kept in a circular buffer, and out-of-order segments (red) are written in the space past the received data.}
        \label{fig:receive_buffer_final}
    \end{subfigure}
    \caption{Na\"ive and final TCP receive buffers}
    \label{fig:receive_buffer}
\end{figure}

We opted for a flat array-based circular buffer for the receive buffer in \sys{}, primarily owing to its determinism in a limited-memory environment: buffer space is reserved at \emph{compile-time}.
Head/tail pointers delimit which part of the array stores in-sequence data. To reduce memory consumption, we store out-of-order data in the same receive buffer, at the same position as if they were in-sequence. We use a bitmap, not head/tail pointers, to record where out-of-order data are stored, because out-of-order data need not be contiguous. We call this an \emph{in-place reassembly queue} (Figure \ref{fig:receive_buffer}).

\section{TCP in a Low-Power Network}\label{sec:interactions}

In this section, we characterize how full-scale TCP interacts with a low-power network stack, resource-constrained hardware, and a low-bandwidth link.

\subsection{Reducing Header Overhead using MSS}\label{sec:mss}

\begin{table}[t]
    \centering
    \setlength\tabcolsep{3.5pt}
    \begin{tabular}{| l || c | c | c | c |}\hline
        & Fast Ethernet & Wi-Fi & Ethernet & 802.15.4\\\hline\hline
        Capacity & 100 Mb/s & 54 Mb/s & 10 Mb/s & 250 kb/s \\\hline
        MTU & 1500 B & 1500 B & 1500 B & 104--116 B \\\hline
        Tx Time & 0.12 ms & 0.22 ms & 1.2 ms & 4.1 ms \\\hline
    \end{tabular}
    \caption{Comparison of TCP/IP links}
    \label{table:link_comparison}
\end{table}

\begin{table}[t]
    \centering
    \setlength\tabcolsep{1pt}
    \begin{tabular}{| c || c | c | c | c || c |} \hline
        Header & 802.15.4 & 6LoWPAN & IPv6 & TCP & Total\\\hline\hline
        1st Frame & 11--23 B & 5 B & 2--28 B & 20--44 B & 38--107 B\\\hline
        $n$th Frame & 11--23 B & 5--12 B & 0 B & 0 B & 16--35 B\\\hline
    \end{tabular}
    \caption{Header overhead with 6LoWPAN fragmentation}
    \label{table:header_overhead}
\end{table}

In traditional networks, it is customary to set the Maximum Segment Size (MSS) to the link MTU (or path MTU) minus the size of the TCP/IP headers.
IEEE 802.15.4 frames, however, are \emph{an order of magnitude smaller} than frames in traditional networks (Table \ref{table:link_comparison}). The TCP/IP headers consume more than half of the frame's available MTU. As a result, TCP performs poorly, incurring more than 50\% header overhead.

Earlier approaches to running TCP over low-MTU links (e.g., low-speed serial links) have used TCP/IP header compression based on per-flow state~\cite{jacobson1990compressing} to reduce header overhead. In contrast, the 6LoWPAN adaptation layer~\cite{montenegro2008lowpan}, designed for LLNs, supports only \emph{flow-independent} compression of the IPv6 header using shared link-layer state, a clear departure from per-flow techniques.
A key reason for this is that the compressor and decompressor in an LLN (host and border router) are separated by several IP hops,\footnote{Thread deliberately does not abstract the mesh as a single IP link. Instead, it organizes the LLN mesh as a set of \emph{overlapping link-local scopes}, using IP-layer routing to determine the path packets take through the mesh~\cite{hui2008ip}.} making it desirable for intermediate nodes to be able to determine a packet's IP header without per-flow context (see \S{}10 of \cite{montenegro2008lowpan}).

That said, compressing TCP headers separately from IP addresses using per-flow state is a promising approach to further amortize header overhead. There is preliminary work in this direction~\cite{ayadi2011implementation, ayadi2010tcp}, but it is based on uIP, which has one in-flight segment, and does not fully specify how to resynchronize compression state after packet loss with a multi-segment window. It is also not officially standardized by the IETF.

Therefore, this paper takes an approach orthogonal to header compression. We instead choose an MSS larger than the link MTU admits, \emph{relying on fragmentation at the lower layers to decrease header overhead}. Fragmentation is handled by 6LoWPAN, which acts at Layer 2.5, between the link and network layers. Unlike end-to-end IP fragmentation, the 6LoWPAN fragments exist only within the LLN, and are reassembled into IPv6 packets when leaving the network.

Relying on fragmentation is effective because, as shown in Table \ref{table:header_overhead}, TCP/IP headers consume space in the first fragment, but not in subsequent fragments. Using an excessively large MSS, however, decreases reliability because the loss of one fragment results in the loss of an entire packet. Existing work~\cite{ayadi2011tcp} has identified this trade-off and investigated it in simulation in the context of power consumption. We investigate it in the context of goodput in a live network.

Figure \ref{fig:mss_bandwidth} shows the bandwidth as the MSS varies.
As expected, we see poor performance at a small MSS due to header overhead.
Performance gains diminish when the MSS becomes larger than 5 frames. We recommend using an MSS of about 5 frames, but it is reasonable to decrease it to 3 frames if more wireless loss is expected. \textbf{Despite the small frame size of IEEE 802.15.4, we can effectively amortize header overhead for TCP using an atypical MSS.} Adjusting the MSS is orthogonal to TCP header compression. We hope that widespread use of TCP over 6LoWPAN, perhaps based on our work, will cause TCP header compression to be separately investigated and possibly used together with a large MSS.

\subsection{Impact of Buffer Size}\label{sec:window}

\begin{figure}[t]
    \centering
    \hspace{-2ex}
    \begin{subfigure}[p]{0.53\linewidth}
        \includegraphics[width=\linewidth]{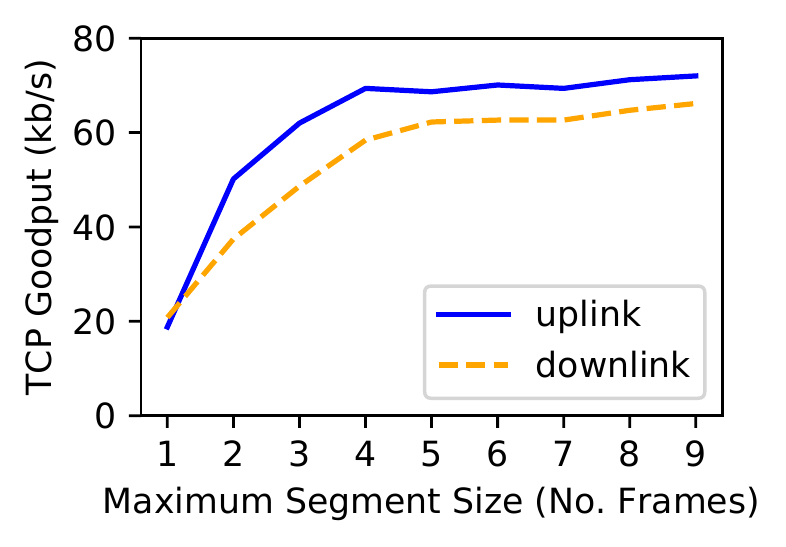}
        \caption{Effect of varying MSS}
        \label{fig:mss_bandwidth}
    \end{subfigure}
    \hspace{-2ex}
    \begin{subfigure}[p]{0.52\linewidth}
        \includegraphics[width=\linewidth]{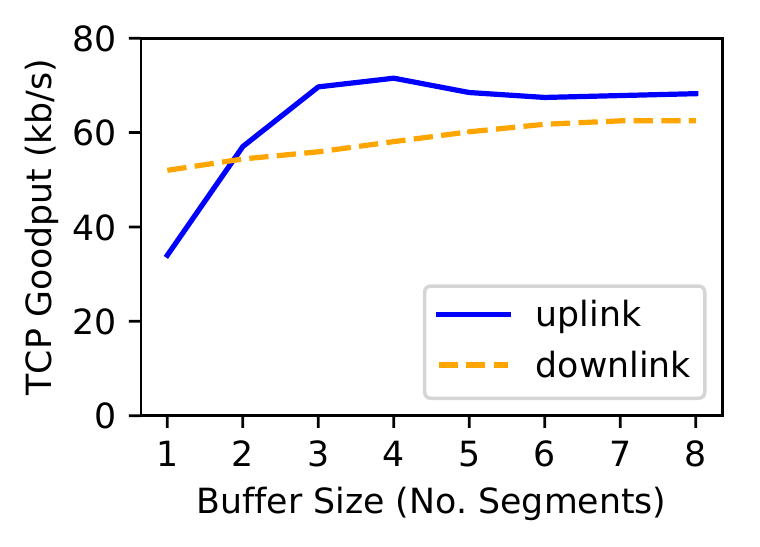}
        \caption{Effect of varying buffer size}
        \label{fig:win_bandwidth}
    \end{subfigure}
    \hspace{-2ex}
    \caption{TCP goodput over one IEEE 802.15.4 hop}
\end{figure}

Whereas simple TCP stacks, like uIP, allow only one in-flight segment, full-scale TCP requires complex buffering (\secref{sec:buffering}). In this section, we vary the size of the buffers (send buffer for uplink experiments and receive buffer for downlink experiments) to study how it affects the bandwidth. In varying the buffer size, we are directly affecting the size of TCP's flow window.
We expect throughput to increase with the flow window size, with diminishing returns once it exceeds the bandwidth-delay product (BDP).
The result is shown in Figure \ref{fig:win_bandwidth}.
\textbf{Goodput levels off at a buffer size of 3 to 4 segments (1386 B to 1848 B), indicating that the buffer size needed to fill the BDP fits comfortably in memory.} Indeed, the BDP in this case is about $125 \text{kb/s} \cdot 0.1 \text{s} \approx 1.6 \text{KiB}$.\footnote{We estimate the bandwidth as $125$ kb/s rather than $250$ kb/s to account for the radio overhead identified in \secref{sec:upper_bound}.}

Downlink goodput at a buffer size of one segment is unusually high. This is because FreeBSD does not delay ACKs if the receive buffer is full, reducing the effective RTT from $\approx 130$ ms to $\approx 70$ ms. Indeed, goodput is very sensitive to RTT when the buffer size is small, because TCP exhibits ``stop-and-wait'' behavior due to the small flow window.

\begin{figure}
    \centering
    \begin{subfigure}[p]{0.445\linewidth}
        \hspace{-1ex}
        \vspace{-0.9ex}
        \includegraphics[width=1.07\linewidth]{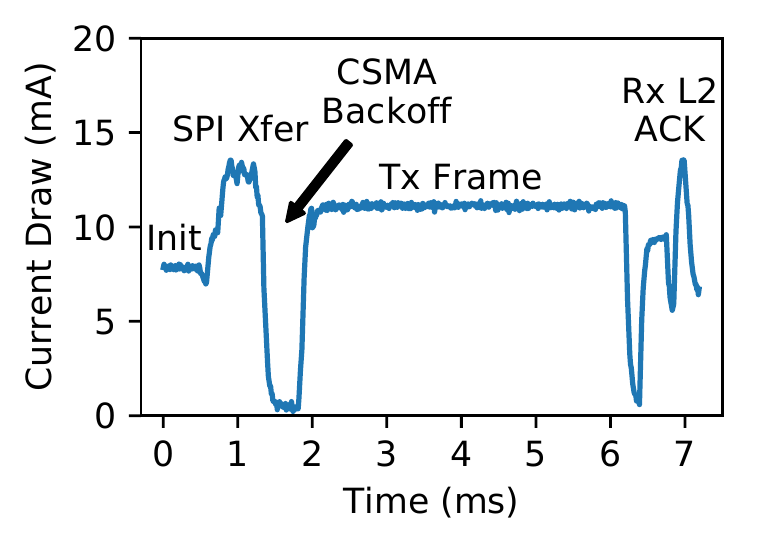}
        \caption{Unicast of a single frame, measured with an oscilloscope}
        \label{fig:current_tx_frame}
    \end{subfigure}
    \hspace{0.05ex}
    \begin{subfigure}[p]{0.535\linewidth}
        \hspace{-1ex}
        \includegraphics[width=1.04\linewidth]{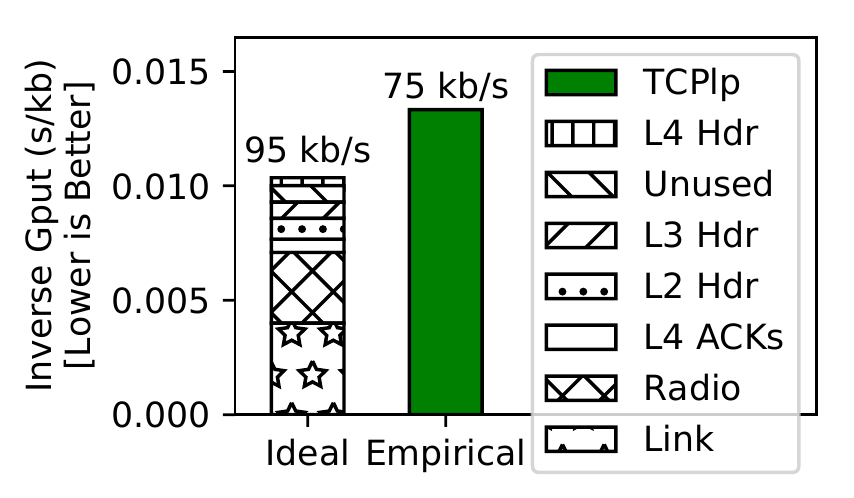}
        \caption{\sys{} goodput compared with raw link bandwidth and overheads}
        \label{fig:breakdown_1hop}
    \end{subfigure}
    \caption{Analysis of overhead limiting \sys{}'s goodput}
    \label{fig:overheads}
\end{figure}

\subsection{Upper Bound on Single-Hop Goodput}\label{sec:upper_bound}

We consider TCP goodput between two nodes over the IEEE 802.15.4 link, over a single hop without any border router.
Using the Hamilton/OpenThread platform, we are able to achieve 75 kb/s.\footnote{\appref{sec:direct} provides the corresponding goodput figures for Hamilton/GNRC and Firestorm/BLIP platforms, for comparison.}
Figure \ref{fig:breakdown_1hop} lists various sources of overhead that limit \sys{}'s goodput, along with the ideal upper bounds that they admit.
\textbf{Link} overhead refers to the 250 kb/s link capacity.
\textbf{Radio} overhead includes SPI transfer to/from the radio (i.e., packet copying~\cite{osterlind2008approaching}), CSMA, and link-layer ACKs, which cannot be pipelined because the AT86RF233 radio has only one frame buffer.
A full-sized 127-byte frame spends 4.1 ms in the air at 250 kb/s, but the radio takes 7.2 ms to send it (Figure \ref{fig:current_tx_frame}), almost halving the link bandwidth available to a single node.
This is consistent with prior results~\cite{osterlind2008approaching}.
\textbf{Unused} refers to unused space in link frames due to inefficiencies in the 6LoWPAN implementation.
Overall, we estimate a 95 kb/s upper bound on goodput (100 kb/s without TCP headers).
Our 75 kb/s measurement is within 25\% of this upper bound, substantially higher than prior work (Table \ref{table:bandwidth_comparison}).
The difference from the upper bound is likely due to network stack processing and other real-world inefficiencies.

\begin{table*}[th]
    \centering
    \setlength\tabcolsep{5pt}
    \begin{tabular}{| l || c | c | c | c | c || c |} \hline
     & \cite{zheng2011tcp} & \cite{ayadi2011implementation} & \cite{hewage2015enabling} &
     \cite{kim2015measurement}\protect\footnotemark{} & \cite{hui2008ip,jonathanhui} & This Paper (Hamilton Platform)\\ \hline\hline
     TCP Stack & uIP & uIP & uIP & BLIP & Arch Rock & \sys{} (RIOT OS, OpenThread)\\ \hline
     Max. Seg Size & 1 Frame & 1 Frame & 4 Frames & 1 Frame & 1024 bytes & 5 Frames\\ \hline
     Window Size & 1 Seg. & 1 Seg. & 1 Seg. & 1 Seg. & 1 Seg. & 1848 bytes (4 Seg.)\\ \hline
     Goodput (One Hop) & 1.5 kb/s & $\approx$ 13 kb/s & $\approx$ 12 kb/s & $\approx$ 4.8 kb/s & 15 kb/s & 75 kb/s\\ \hline
     Goodput (Multi-Hop) & $\approx$ 0.55 kb/s & $\approx$ 6.4 kb/s & $\approx$ 12 kb/s & $\approx$ 2.4 kb/s & 9.6 kb/s & 20 kb/s \\ \hline
    \end{tabular}
    \caption{Comparison of \sys{} to existing TCP implementations used in network studies over IEEE 802.15.4 networks.\protect\footnotemark{} Goodput figures obtained by reading graphs in the original paper (rather than stated numbers) are marked with the $\approx$ symbol.}
    \label{table:bandwidth_comparison}
\end{table*}
\addtocounter{footnote}{-1}
\footnotetext{For this study, we list aggregate goodput over multiple TCP flows.}
\stepcounter{footnote}
\footnotetext{One study~\cite{duquennoy2011lossy} achieves $\approx$ 16 kb/s over multiple hops using the Linux TCP stack. We do not include it in Table \ref{table:bandwidth_comparison} because it does not capture the resource constraints of LLNs (it uses traditional computers for the end hosts) and does not consider hidden terminals (it uses different wireless channels for different wireless hops). It uses TCP to evaluate link-layer burst forwarding.}

\begin{figure*}[t]
    \centering
    \hspace{-2ex}
    \begin{subfigure}[p]{0.26\linewidth}
        \includegraphics[width=\linewidth]{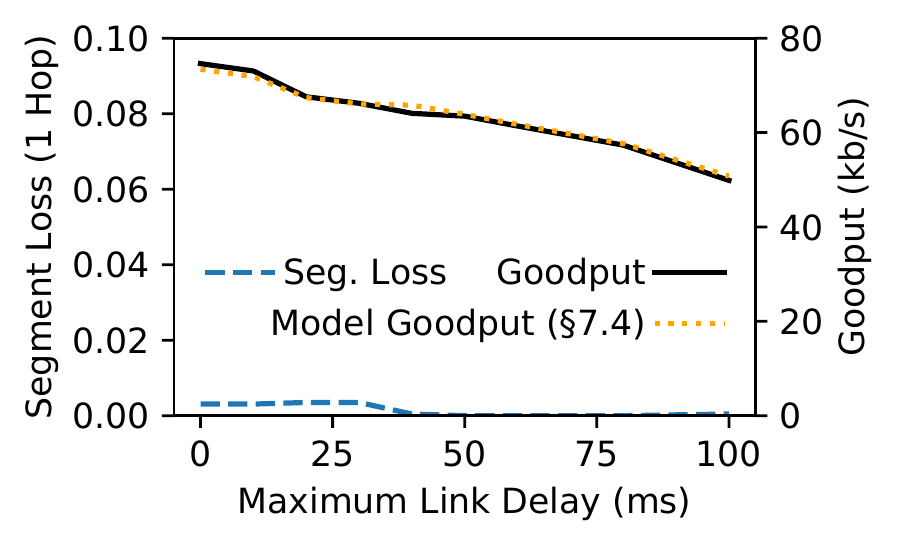}
        \caption{TCP goodput, one hop}
        \label{fig:lld_1hop}
    \end{subfigure}
    \hspace{-1.5ex}
    \begin{subfigure}[p]{0.26\linewidth}
        \includegraphics[width=\linewidth]{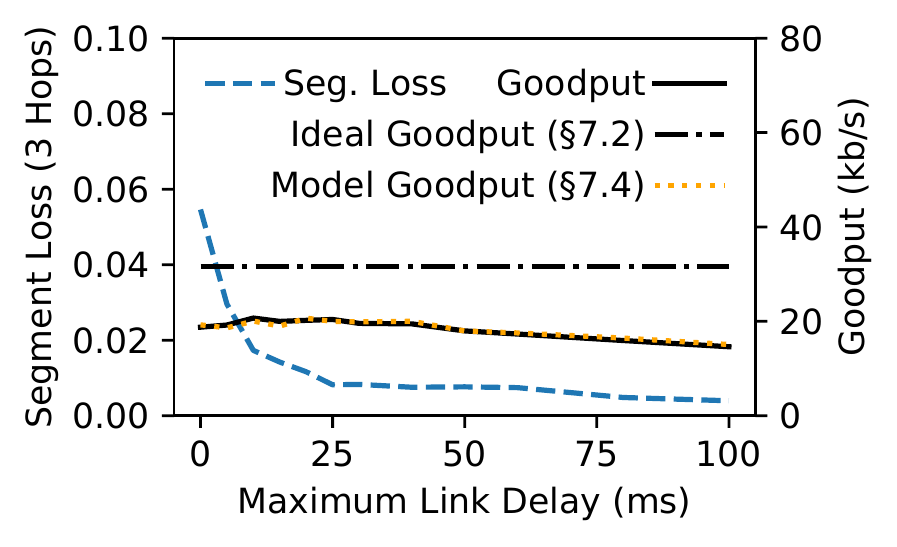}
        \caption{TCP goodput, three hops}
        \label{fig:lld_3hops}
    \end{subfigure}
    \hspace{-1.5ex}
    \begin{subfigure}[p]{0.28\linewidth}
        \includegraphics[width=\linewidth]{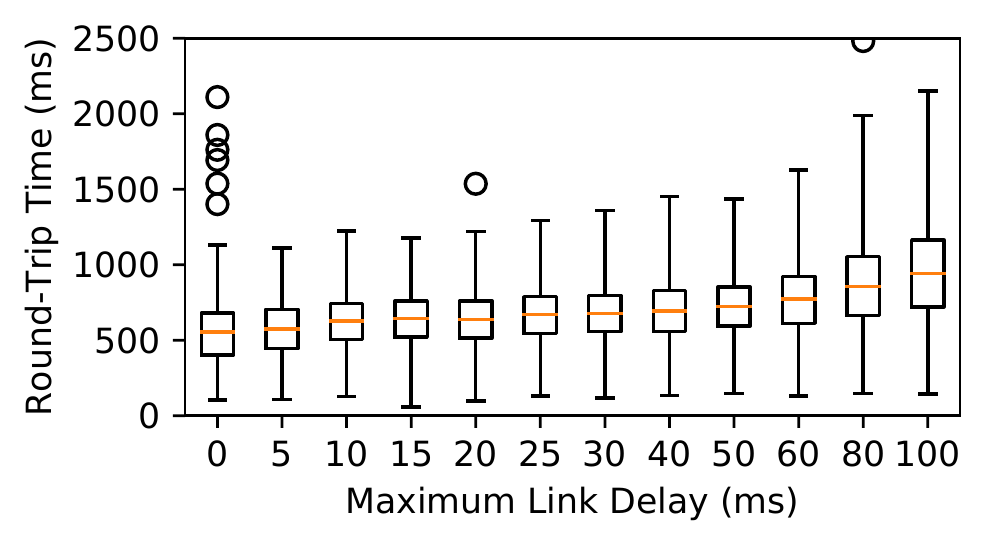}
        \caption{RTT, three hops (outliers omitted)}
        \label{fig:lld_3hops_rtt}
    \end{subfigure}
    \hspace{-1.5ex}
    \begin{subfigure}[p]{0.22\linewidth}
        \centering
        \includegraphics[width=0.9\linewidth]{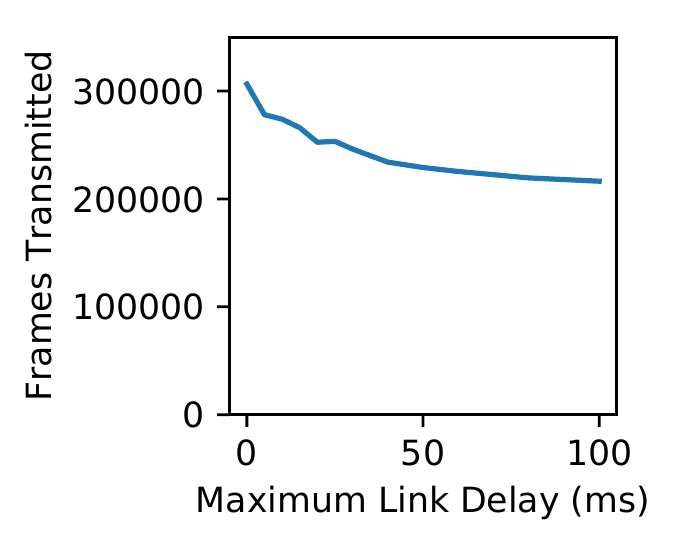}
        \caption{Total frames sent, three hops}
        \label{fig:lld_3hops_frames}
    \end{subfigure}
    \hspace{-3ex}
    \caption{Effect of varying time between link-layer retransmissions. Reported ``segment loss'' is the loss rate of TCP segments, not individual IEEE 802.15.4 frames. It includes only losses not masked by link-layer retries.}
    \label{fig:lld}
    \vspace{-1ex}
\end{figure*}

\section{TCP Over Multiple Wireless Hops}\label{sec:multihop}

We instrument TCP connections between Hamilton nodes in our multi-hop testbed, without using the EC2 server.

\subsection{Mitigating Hidden Terminals in LLNs}\label{ssec:lldelay}

Prior work over traditional WLANs has shown that hidden terminals degrade TCP performance over multiple wireless hops~\cite{gerla1999tcp}.
Using RTS/CTS for hidden terminal avoidance has been shown to be effective in WLANs. This technique has an unacceptably high overhead in LLNs~\cite{woo2001transmission}, however, because data frames are small (Table \ref{table:link_comparison}), comparable in size to the additional control frames required.
Prior work in LLNs has carefully designed application traffic, using rate control~\cite{kim2007flush, hull2004mitigating} and link-layer delays~\cite{woo2001transmission}, to avoid hidden terminals.

But prior work does not explore these techniques in the context of TCP.
Unlike protocols like CoAP and simplified TCP implementations like uIP, a full-scale TCP flow has a \emph{multi-segment sliding window} of unacknowledged data, making it unclear \emph{a priori} whether existing LLN techniques will be sufficient. In particular, rate control seems sufficient because of bi-directional packet flow in TCP (data in one direction and ACKs in the other).
So, rather than applying rate control, we attempt to avoid hidden terminals by adding a delay $d$ between link-layer retries in addition to CSMA backoff.
After a failed link transmission, a node waits for a random duration between $0$ and $d$, before retransmitting the frame. The idea is that if two frames collide due to a hidden terminal, the delay will prevent their link-layer retransmissions from colliding.

We modified OpenThread, which previously had no delay between link retries, to implement this.
As expected, single-hop performance (Figure \ref{fig:lld_1hop}) decreases as the delay between link retries increases; hidden terminals are not an issue in that setting.
Packet loss is high for the multihop experiment (Figure \ref{fig:lld_3hops}) when the link retry delay is 0, as is expected from hidden terminals. \textbf{Adding a small delay between link retries, however, effectively reduces packet loss.} Making the delay too large raises the RTT (Figure \ref{fig:lld_3hops_rtt}).

We prefer a smaller frame/segment loss rate, even if goodput stays the same, in order to make more efficient use of network resources. Therefore, we prefer a moderate delay ($d = 40$ ms) to a small delay ($d = 5$ ms), even though both provide the same goodput, because the frame and segment loss rates are smaller when $d$ is large (Figures \ref{fig:lld_3hops} and \ref{fig:lld_3hops_frames}).

\subsection{Upper Bound on Multi-Hop Goodput}

Comparing Figures \ref{fig:lld_1hop} and \ref{fig:lld_3hops}, goodput over three wireless hops is substantially smaller than goodput over a single hop. Prior work has observed similar throughput reductions over multiple hops~\cite{kim2015measurement, osterlind2008approaching}. It is due to radio scheduling constraints inherent in the multihop setting, which we describe in this section.
Let $B$ be the bandwidth over a single hop.

Consider a two-hop setup: $S \rightarrow R_1 \rightarrow D$. $R_1$ cannot receive a frame from $S$ while sending a frame to $D$, because its radio cannot transmit and receive simultaneously. Thus, the maximum achievable bandwidth over two hops is $\frac{B}{2}$.

Now consider a three-hop setup: $S \rightarrow R_1 \rightarrow R_2 \rightarrow D$. By the same argument, if a frame is being transferred over $R_1 \rightarrow R_2$, then neither $S \rightarrow R_1$ nor $R_2 \rightarrow D$ can be active. Furthermore, if a frame is being transferred over $R_2 \rightarrow D$, then $R_1$ can hear that frame. Therefore, $S \rightarrow R_1$ cannot transfer a frame at that time; if it does, then its frame will collide at $R_1$ with the frame being transferred over $R_2 \rightarrow D$.
Thus, the maximum bandwidth is $\frac{B}{3}$.
We depict this ideal upper bound in Figure \ref{fig:lld_3hops}, taking $B$ to be the ideal single-hop goodput from \secref{sec:upper_bound}.

In setups with more than three hops, every set of three adjacent hops is subject to this constraint. The first hop and fourth hop, however, may be able to transfer frames simultaneously. Therefore, the maximum bandwidth is still $\frac{B}{3}$. In practice, goodput may fall slightly because transmissions from a node may \emph{interfere} with nodes multiple hops away, even if they can only be received by its immediate neighbors.

We made empirical measurements with $d = 40$ ms to validate this analysis.
Goodput over one hop was 64.1 kb/s; over two hops, 28.3 kb/s; over three hops, 19.5 kb/s; and over four hops, 17.5 kb/s. This roughly fits the model.

This analysis justifies why the same window size works well for both the one-hop experiments and the three-hop experiments in \secref{ssec:lldelay}. Although the RTT is three times higher, the bandwidth-delay product is approximately the same. \textbf{Crucially, this means that the 2 KiB buffer size we determined in \secref{sec:window}, which fits comfortably in memory, remains applicable for up to three wireless hops.}

\subsection{TCP Congestion Control in LLNs}\label{ssec:congestion}

Recall that small send/receive buffers of only 1848 bytes (4 TCP segments) each are enough to achieve good TCP performance. This profoundly impacts TCP's congestion control mechanism. For example, consider Figure \ref{fig:lld_3hops}. It is remarkable that throughput is almost the same at $d = 0$ ms and $d = 30$ ms, despite having 6\% packet loss in the first case and less than 1\% packet loss in the second.

\begin{figure}[t]
    \centering
    \hspace{-3ex}
    \begin{subfigure}[p]{0.53\linewidth}
        \includegraphics[width=\linewidth]{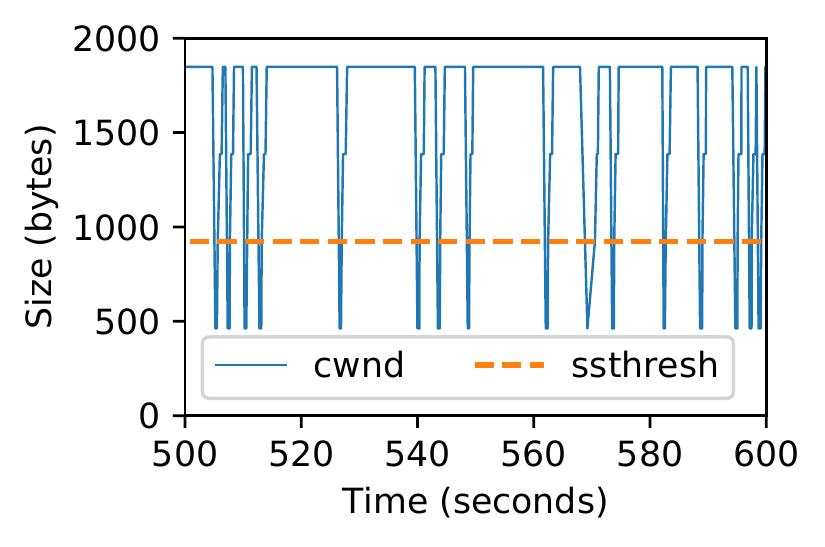}
        \caption{TCP \cwnd{} for $d = 0$, three hops}
        \label{fig:lld_3hops_cwnd}
    \end{subfigure}
    \hspace{-1ex}
    \begin{subfigure}[p]{0.51\linewidth}
        \includegraphics[width=\linewidth]{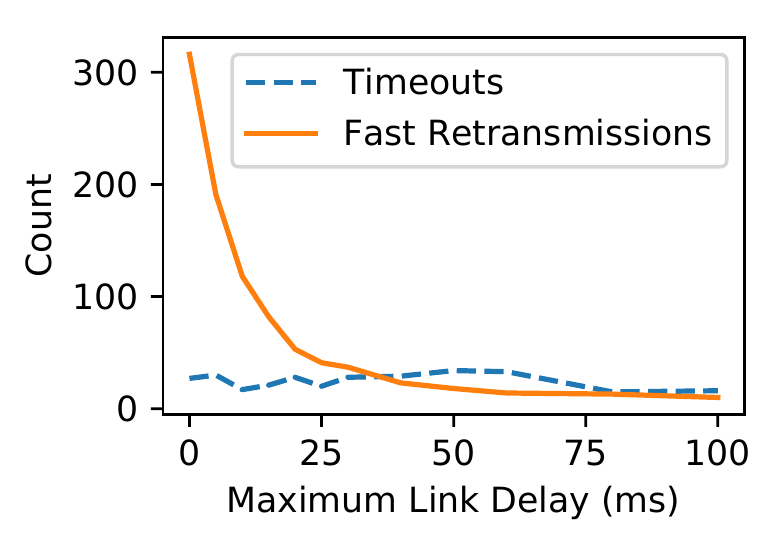}
        \caption{TCP loss recovery, three hops}
        \label{fig:lld_3hops_losses}
    \end{subfigure}
    \hspace{-3ex}
    \caption{Congestion behavior of TCP over IEEE 802.15.4}
    \label{fig:congestion}
\end{figure}

Figure \ref{fig:lld_3hops_cwnd} depicts the congestion window over a 100 second interval during the $d = 0$ ms experiment.\footnote{All congestion events in Figure \ref{fig:lld_3hops_cwnd} were fast retransmissions, except for one timeout at $t = 569$ s. \cwnd{} is temporarily set to $1$ MSS during fast retransmissions due to an artifact of FreeBSD's implementation of SACK recovery. For clarity, we cap \cwnd{} at the size of the send buffer, and we remove fluctuations in \cwnd{} which resulted from ``bad retransmissions'' that the FreeBSD implementation corrected in the course of its normal execution.}
Interestingly, the \cwnd{} graph is far from the canonical sawtooth shape (e.g., Figure 11(b) in \cite{balakrishnan1997comparison}); \cwnd{} is almost always maxed out even though losses are frequent (6\%).
This is specific to small buffers. In traditional environments, where links have higher throughput and buffers are large, it takes longer for \cwnd{} to recover after packet loss, greatly limiting the sending rate with frequent packet losses. In contrast, \textbf{in LLNs, where send/receive buffers are small, \cwnd{} recovers to the maximum size quickly after packet loss, making TCP performance robust to packet loss.}

Congestion behavior also provides insight into loss patterns, as shown in Figure \ref{fig:lld_3hops_losses}. Fast retransmissions (used for isolated losses) become less frequent as $d$ increases, suggesting that they are primarily caused by hidden-terminal-related losses. Timeouts do not become less frequent as $d$ is increased, suggesting that they are caused by something else.

\subsection{Modeling TCP Goodput in an LLN}\label{sec:model}

Our findings in \secref{ssec:congestion} suggest that, in LLNs, \cwnd{} is limited by the buffer size, not packet loss.
To validate this, we analytically model TCP performance according to our observations in \secref{ssec:congestion}, and then check if the resulting model is consistent with the data.
Comprehensive models of TCP, which take window size limitations into account, already exist~\cite{padhye1998modeling}; in contrast, our model is \emph{intentionally simple} to provide intuition.

Observations in \secref{ssec:congestion} suggest that we can neglect the time it takes the congestion window to recover after packet loss.
So, we model a TCP connection as \textit{binary}: either it is sending data with a full window, or it is not sending new data because it is recovering from packet loss.
According to this model, a TCP flow alternates between \emph{bursts} when it is transmitting at a full window, and \emph{rests} when it is in recovery and not sending new data. Burst lengths depend on the packet loss rate $p$ and rest lengths depend on RTT.
This approach leads to the following model (full derivation is in Appendix \ref{sec:derivation}):
\begin{equation}\label{eq:lln_model}
    B = \frac{\mss}{\rtt} \cdot \frac{1}{\frac{1}{w} + 2p}
\end{equation}
\noindent
where $B$, the TCP goodput, is written in terms of the maximum segment size $\text{MSS}$, round-trip time $\text{RTT}$, packet loss rate $p$ ($0<p<1$),
and window size $w$ (sized to BDP, in packets).
Figures \ref{fig:lld_1hop} and \ref{fig:lld_3hops} include the predicted goodput as dotted lines, calculated according to Equation \ref{eq:lln_model} using the empirical RTT and segment loss rate for each experiment. \textbf{Our model of TCP goodput closely matches the empirical results.}

An established model of TCP outside of LLNs is~\cite{kurose2013model, mathis1997macroscopic}:
\begin{equation}\label{eq:newreno_model}
B = \frac{\text{MSS}}{\text{RTT}} \cdot \sqrt{\frac{3}{2p}}
\end{equation}
\noindent
Equation \ref{eq:newreno_model} fundamentally relies on there being many competing flows, so we do not expect it to match our empirical results from \secref{ssec:congestion}. But, given that existing work examining TCP in LLNs makes use of this formula to ground new algorithms~\cite{im2015tcp}, the differences between Equations \ref{eq:lln_model} and \ref{eq:newreno_model} are interesting to study.
In particular, Equation \ref{eq:lln_model} has an added $\frac{1}{w}$ in the denominator and depends on $p$ rather than $\sqrt{p}$, explaining, mathematically, how TCP in LLNs is more robust to small amounts of packet loss.
We hope Equation \ref{eq:lln_model}, together with Equation \ref{eq:lln_full_model} in \appref{sec:derivation}, will provide a foundation for future research on TCP in LLNs.

\section{TCP in LLN Applications}\label{sec:application}

To demonstrate that TCP is practical for real IoT use cases, we compare its performance to that of CoAP, CoCoA, and unreliable UDP in three workloads inspired by real application scenarios: web server, sense-and-send, and event detection. We evaluate the protocols over multiple hops with duty-cycled radios and wireless interference, present in our testbed in the day (\secref{s:hardware}). In our experiments, nodes 12--15 (Figure \ref{fig:topology}) send data to a server running on Amazon EC2. The RTT from the border router to the server was $\approx 12$ ms, much smaller than within the low-power mesh ($\approx 100$-$300$ ms).

In our preliminary experiments, we found that in the presence of simultaneous TCP flows, tail drops at a relay node significantly impacted fairness. Implementing Random Early Detection (RED)~\cite{floyd1993random} with Explicit Congestion Notification (ECN) support solved this problem. Therefore, we use RED and ECN for experiments in this section with multiple flows. While such solutions have sometimes been problematic since they are implemented in routers, they are more natural in LLNs because the intermediate ``routers'' relaying packets in an LLN typically also participate in the network as hosts.

We generally use a smaller MSS (3 frames) in this section, because it is more robust to interference in the day (\secref{sec:interactions}).
We briefly discuss how this affects our model in \appref{sec:derivation}, but leave a rigorous treatment to future work.

Running TCP in these application scenarios motivates \textbf{Adaptive Duty Cycle} and \textbf{Finer-Grained Link Queue Management}, which we introduce below as they are needed.

\subsection{Web Server Application Scenario}\label{ssec:webserver}

To study TCP with multiple wireless hops and duty cycling, we begin with a web server hosted on a low-power device. We compare HTTP/TCP and CoAP/UDP (\secref{sec:thread}).

\begin{figure}[t]
    \centering
    \hspace{-2ex}
    \begin{subfigure}[p]{0.35\linewidth}
        \includegraphics[width=\linewidth]{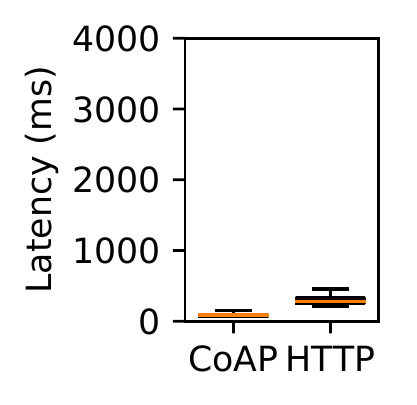}
        \caption{No duty cycling}
        \label{fig:webserver_latency_nodutycycle}
    \end{subfigure}
    \hspace{-2ex}
    \begin{subfigure}[p]{0.35\linewidth}
        \includegraphics[width=\linewidth]{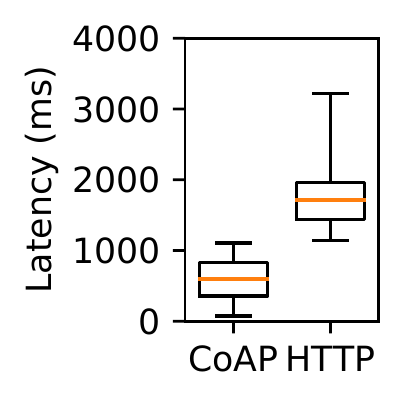}
        \caption{1 s sleep interval}
        \label{fig:webserver_latency_1s_notadaptive}
    \end{subfigure}
    \hspace{-2ex}
    \begin{subfigure}[p]{0.35\linewidth}
        \includegraphics[width=\linewidth]{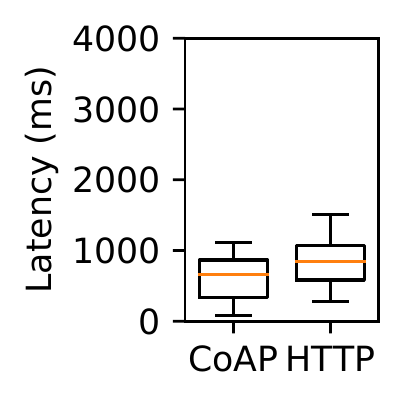}
        \caption{1 s sleep interval with adaptive duty cycle}
        \label{fig:webserver_latency_1s_adaptive}
    \end{subfigure}
    \hspace{-2ex}
    \caption{Latency of web request: CoAP vs. HTTP/TCP}
    \label{fig:webserver_latency}
\end{figure}

\subsubsection{Latency Analysis}\label{ssec:webserver_latency}
An HTTP request requires two round-trips: one to establish a TCP connection, and another for request/response. CoAP requires only one round trip (no connection establishment) and has smaller headers. Therefore, CoAP has a lower latency than HTTP/TCP when using an always-on link (Figure \ref{fig:webserver_latency_nodutycycle}). Even so, the latency of HTTP/TCP in this case is well below 1 second, not so large as to degrade user experience.

We now explore how a duty-cycled link affects the latency. Recall that leaf nodes in OpenThread (\secref{sec:thread}) periodically poll their parent to receive downstream packets, and keep their radios in a low-power sleep state between polls. We set the \emph{sleep interval}---the time that a node waits between polls---to 1 s and show the latency in Figure \ref{fig:webserver_latency_1s_notadaptive}. Interestingly, HTTP's minimum observed latency is much higher than CoAP's, more than is explained by its additional round trip.

Upon investigation, we found that this is because \textbf{the self-clocking nature of TCP~\cite{jacobson1988congestion} interacts poorly with the duty-cycled link}. Concretely, the web server receives the SYN packet when it polls its parent, and sends the SYN-ACK immediately afterward, at the \emph{beginning} of the next sleep interval. The web server therefore waits for the \emph{entire} sleep interval before polling its parent again to receive the HTTP request, thereby experiencing the worst-case latency for the second round trip. We also observed this problem for batch transfer over TCP; TCP's self-clocking behavior causes it to consistently experience the worst-case round-trip time.

To solve this problem, we propose a technique called \textbf{Adaptive Duty Cycling}. After the web server receives a SYN, it \emph{reduces the sleep interval} in anticipation of receiving an HTTP request. After serving the request, it restores the sleep interval to its old value. Unlike early LLN link-layer protocols like S-MAC~\cite{ye2004medium} that use an adaptive duty cycle, we use \emph{transport-layer state} to inform the duty cycle. Figure \ref{fig:webserver_latency_1s_adaptive} shows the latency with adaptive duty cycling, where the sleep interval is temporarily reduced to 100 ms after connection establishment. \textbf{With adaptive duty-cycling, the latency overhead of HTTP compared to CoAP is small, despite larger headers and an extra round trip for connection establishment.}

Adaptive duty cycling is also useful in high-throughput scenarios, and in situations with persistent TCP connections. We apply adaptive duty cycling to one such scenario in \secref{sec:senseandsend}.

\begin{figure}[t]
    \centering
    \hspace{-3ex}
    \begin{subfigure}[p]{0.54\linewidth}
        \includegraphics[width=\linewidth]{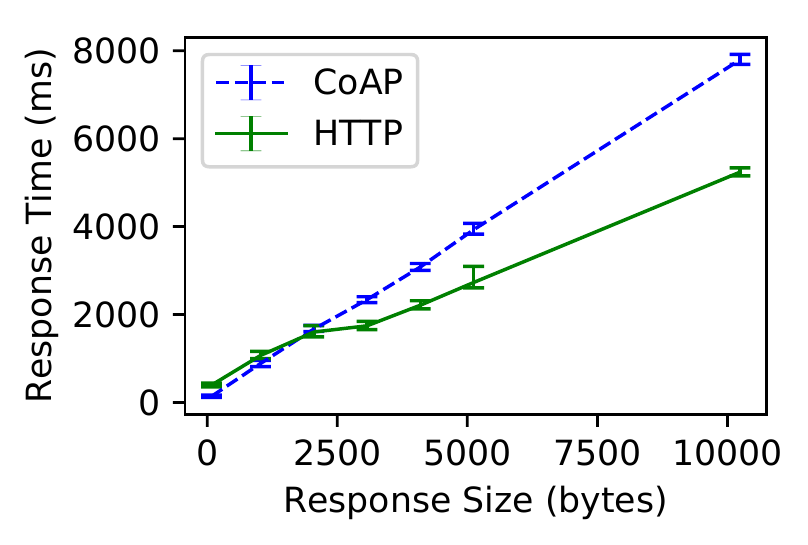}
        \caption{Response time vs. size}
        \label{fig:webserver_throughput_varying}
    \end{subfigure}
    \hspace{-2ex}
    \begin{subfigure}[p]{0.50\linewidth}
        \includegraphics[width=\linewidth]{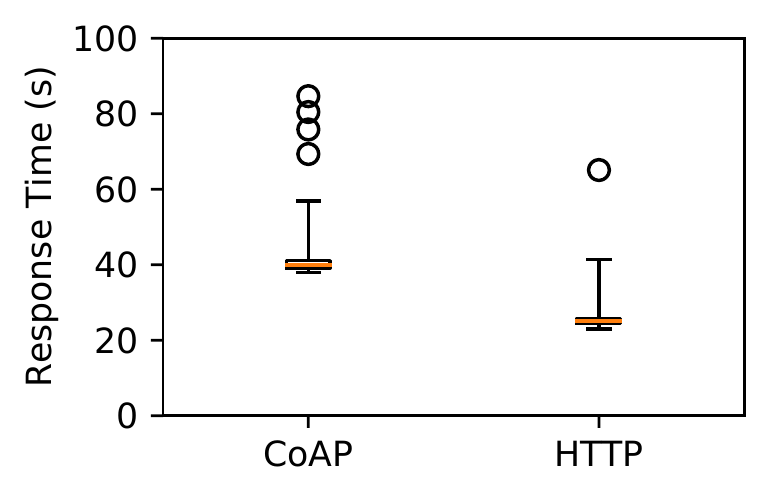}
        \caption{50 KiB response size}
        \label{fig:webserver_throughput_50KiB}
    \end{subfigure}
    \hspace{-4ex}
    \caption{Goodput: CoAP vs. HTTP/TCP}
    \label{fig:webserver_throughput}
\end{figure}

\subsubsection{Throughput Analysis}
In \secref{ssec:webserver_latency}, the size of the web server's response was 82 bytes, intentionally small to focus on latency. In a real application, however, the response may be large (e.g., it may contain a batch of sensor readings). In this section, we explore larger response sizes. We use a short sleep interval of 100 ms. This is realistic because, using adaptive duty cycling, the sleep interval may be longer when the node is idle, and reduced to 100 ms only when transferring the response.

Figure \ref{fig:webserver_throughput_varying} shows the total time from dispatching the request to receiving the full response, as we vary the size of the response. It plots the median time, with quartiles shown in error bars. HTTP takes longer than CoAP when the response size is small (consistent with Figure \ref{fig:webserver_latency}), but CoAP takes longer when the response size is larger. This indicates that while HTTP/TCP has a greater fixed-size overhead than CoAP (higher y-intercept), it transfers data at a higher throughput (lower slope). TCP achieves a higher throughput than CoAP because CoAP sends response segments one-at-a-time (``stop and wait''), whereas TCP allows multiple segments to be in flight simultaneously (``sliding window'').

To quantify the difference in throughput, we compare TCP and CoAP when transferring 50 KiB of data in Figure \ref{fig:webserver_throughput_50KiB}. \textbf{TCP achieves 40\% higher throughput compared to CoAP, over multiple hops and a duty-cycled link.}

\subsubsection{Power Consumption}
TCP consumes more energy than CoAP due to the extra round-trip at the beginning. In practice, however, a web server is interactive, and therefore will be \emph{idle} most of the time. Thus, the idle power consumption dominates. For example, TCP keeps the radio on 35\% longer than CoAP for a response size of 1024 bytes, but if the user makes one request every 100 seconds on average, this difference drops to only 0.35\%.

Thus, we relegate in-depth power measurements to the sense-and-send application (\secref{sec:senseandsend}), which is non-interactive.

\subsection{Sense-and-Send Application Scenario}\label{sec:senseandsend}

We turn our focus to the common \emph{sense-and-send} paradigm, in which  devices periodically collect sensor readings and send them upstream. For concreteness, we model our experiments on the deployment of anemometers in a building, a real-world LLN use case described in \appref{app:anemometer}. Anemometers collect measurements frequently (once per second), making heavy use of the transport protocol; given that our focus is on transport performance, this makes anemometers a good fit for our study.
Other sensor deployments (e.g., temperature, humidity, building occupancy, etc.) sample data at a lower rate (e.g., 0.05 Hz), but are otherwise similar. Thus, \emph{we expect our results to generalize to other sense-and-send applications.}

Nodes 12--15 (Figure \ref{fig:topology}) each generate one 82-byte reading every 1 second, and send it to the cloud server using either TCP or CoAP.
We use most of the remaining RAM as an \emph{application-layer queue} to prevent data from being lost if CoAP or TCP is in backoff after packet loss and cannot send out new data immediately.
We make use of adaptive duty cycling for both TCP and CoAP, with a base sleep interval of four minutes (OpenThread's default) and decreasing it to 100 ms\footnote{100 ms is comparable to ContikiMAC's default sleep interval of 125 ms.} when a TCP ACK or CoAP response is expected.

We measure a solution's \emph{reliability} as the proportion of generated readings delivered to the server. Given that TCP and CoAP both guarantee reliability, a reliability measurement of less than 100\% is caused by overflow of the application-layer queue due to poor network conditions preventing data from being reliably communicated as fast as they are generated. Generating data more slowly would result in higher reliability.

\subsubsection{Performance in Favorable Conditions}
We begin with experiments in our testbed at night, when there is less wireless interference. We compare three setups: (1) CoAP, (2) CoCoA, and (3) \sys{}.
We also compare two sending scenarios: (1) sending each sensor reading right away (``No Batching''), and (2) sending sensor readings in batches of 64 (``Batching'')~\cite{kim2007health}.
We ensure that packets in a CoAP batch are the same size as segments in TCP (five frames).

\begin{figure}[t]
    \centering
    \hspace{-3ex}
    \begin{subfigure}[p]{0.52\linewidth}
        \includegraphics[width=\linewidth]{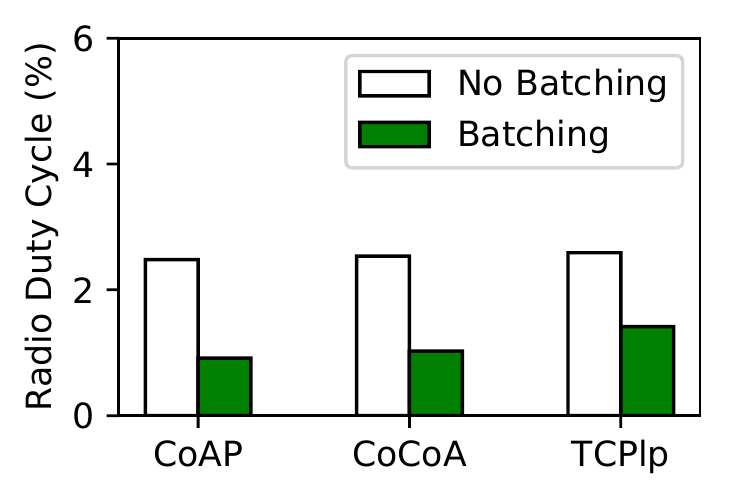}
        \caption{Radio duty cycle}
        \label{fig:rdc_batching}
    \end{subfigure}
    \hspace{-1ex}
    \begin{subfigure}[p]{0.52\linewidth}
        \includegraphics[width=\linewidth]{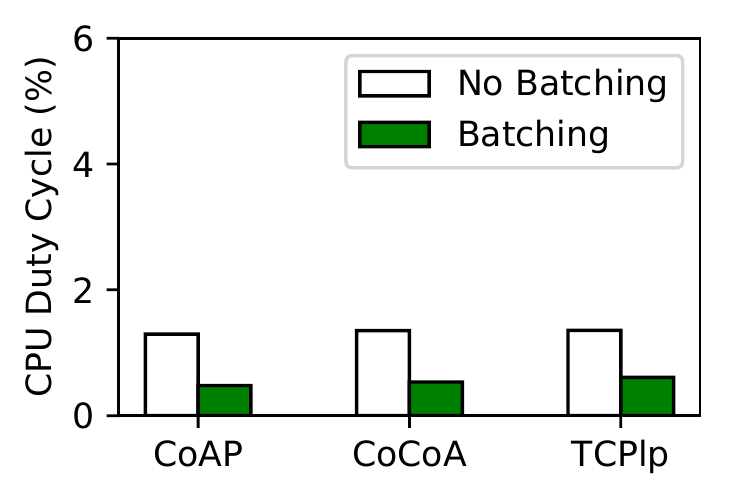}
        \caption{CPU duty cycle}
        \label{fig:cdc_batching}
    \end{subfigure}
    \hspace{-4ex}
    \caption{Effect of batching on power consumption}
    \label{fig:batching}
\end{figure}

All setups achieved 100\% reliability due to end-to-end acknowledgments (figures are omitted for brevity). Figures \ref{fig:rdc_batching} and \ref{fig:cdc_batching} also show that all the three protocols consume similar power; \textit{TCP is comparable to LLN-specific solutions}.

\textbf{Both the radio and CPU duty cycle are significantly smaller with batching than without batching.}
By sending data in batches, nodes can amortize the cost of sending data and waiting for a response.
Thus, batching is the more realistic workload, so we use it to continue our evaluation.

\subsubsection{Resilience to Packet Loss}\label{sec:injected}

\begin{figure}[t]
    \centering
    \hspace{-3ex}
    \begin{subfigure}[p]{0.53\linewidth}
        \includegraphics[width=\linewidth]{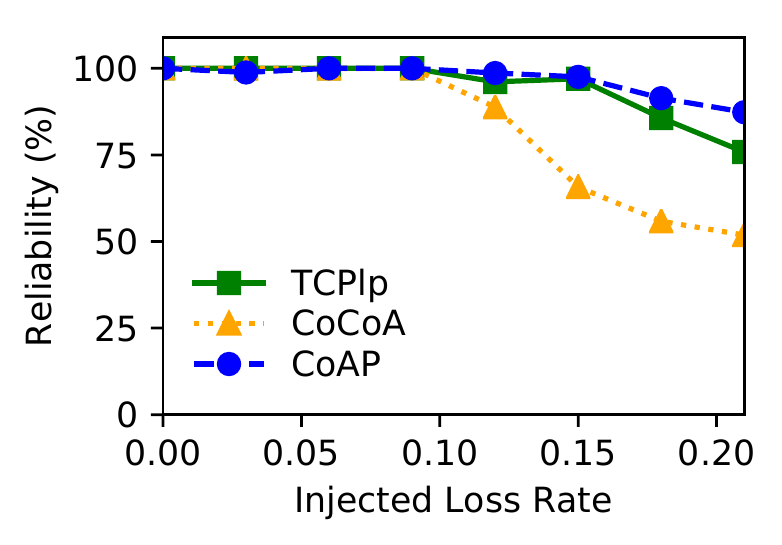}
        \caption{Reliability}
        \label{fig:reliability_injected_loss}
    \end{subfigure}
    \hspace{-1ex}
    \begin{subfigure}[p]{0.52\linewidth}
        \includegraphics[width=\linewidth]{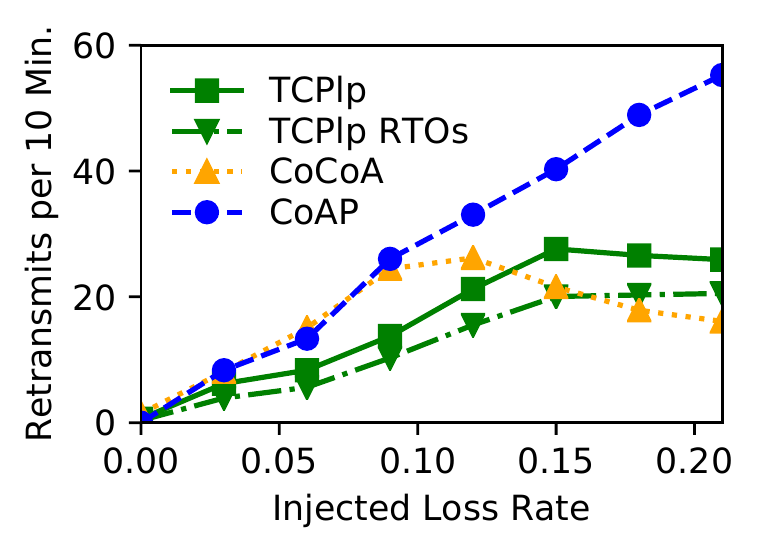}
        \caption{Transport-layer retries}
        \label{fig:rexmit_injected_loss}
    \end{subfigure}
    \hspace{-3ex}

    \hspace{-3ex}
    \begin{subfigure}[p]{0.52\linewidth}
        \includegraphics[width=\linewidth]{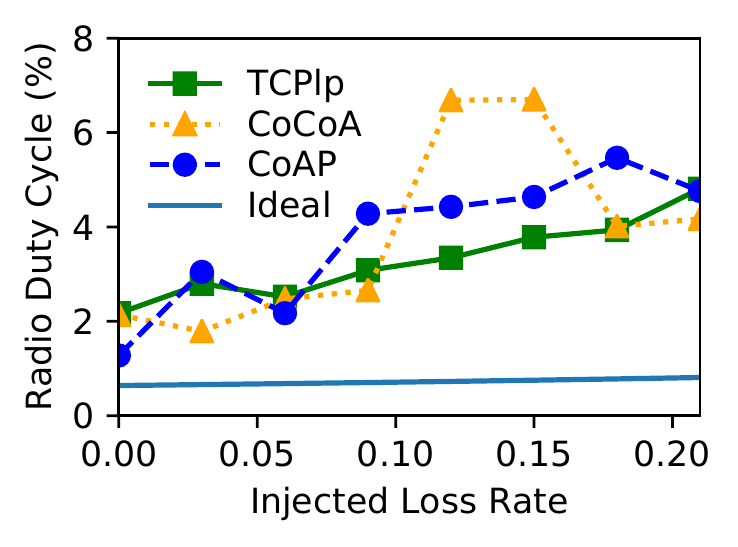}
        \caption{Radio duty cycle}
        \label{fig:rdc_injected_loss}
    \end{subfigure}
    \hspace{-1ex}
    \begin{subfigure}[p]{0.52\linewidth}
        \includegraphics[width=\linewidth]{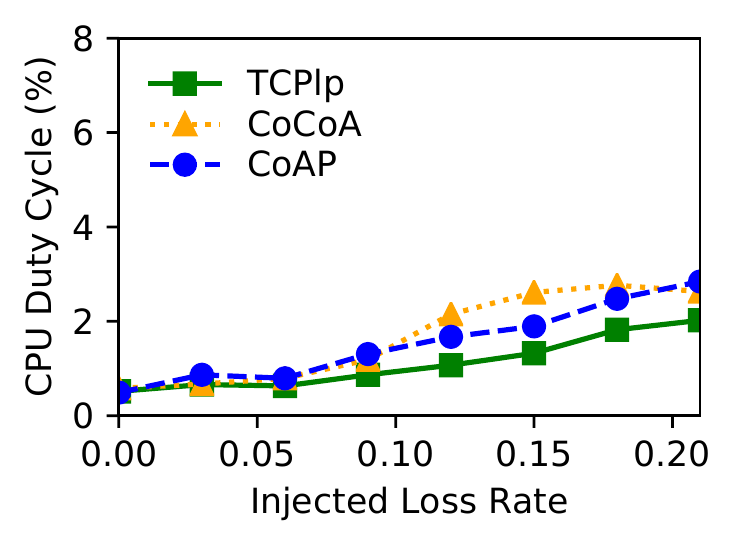}
        \caption{CPU duty cycle}
        \label{fig:cdc_injected_loss}
    \end{subfigure}
    \hspace{-3ex}
    \caption{Performance with injected packet loss}
    \label{fig:injected_loss}
\end{figure}

In this section, we inject uniformly random packet loss at the border router and measure each solution.
The result is shown in Figure \ref{fig:injected_loss}. Note that the injected loss rate corresponds to the \emph{packet-level} loss rate \emph{after} link retries and 6LoWPAN reassembly.
Although we plot loss rates up to 21\%, \emph{we consider loss rates $>15$\% exceptional; we focus on the loss rate up to 15\%}. A number of WSN studies have already achieved $>90$\% end-to-end packet delivery, using only link/routing layer techniques (not transport)~\cite{duquennoy2015orchestra,kim2017dt,kim2015marketnet}. In our testbed environment, we have not observed the loss rate exceed 15\% for an extended time, even with wireless interference.

\textbf{Both CoAP and TCP achieve nearly 100\% reliability} at packet loss rates less than 15\%, as shown in Figure \ref{fig:reliability_injected_loss}. At loss rates greater than 9\%, CoCoA performs poorly. The reason is that CoCoA attempts to measure RTT for retransmitted packets, and conservatively calculates the RTT relative to the first transmission. This results in an inflated RTT value that causes CoCoA to delay longer before retransmitting, causing the application-layer queue to overflow.
Full-scale TCP is immune to this problem despite measuring the RTT, because the TCP timestamp option allows TCP to unambiguously determine the RTT even for retransmitted segments.

Figures~\ref{fig:rdc_injected_loss} and \ref{fig:cdc_injected_loss} show that, overall, \textbf{TCP and CoAP perform comparably in terms of radio and CPU duty cycle}.
At 0\% injected loss, \sys{} has a slightly higher duty cycle, consistent with Figure \ref{fig:batching}.
At moderate packet loss, \sys{} appears to have a slightly lower duty cycle.
This may be due to TCP's sliding window, which allows it to tolerate some ACK losses without retries. Additionally, Figure \ref{fig:rexmit_injected_loss} shows that, although most of TCP's retransmissions are explained by timeouts, a significant portion were triggered in other ways (e.g., duplicate ACKs). In contrast, CoAP and CoCoA rely exclusively on timeouts, which has intrinsic limitations~\cite{zhang1986why}.

With exceptionally high packet loss rates ($>$15\%),
CoAP achieves higher reliability than TCP, because it ``gives up'' after just 4 retries; it exponentially increases the wait time between those retries, but then resets its RTO to 3 seconds when giving up and moving to the next packet.
In contrast, TCP performs up to 12 retries with exponential backoff.
Thus, TCP backs off further than CoAP upon consecutive packet losses, witnessed by the smaller retransmission count in Figure \ref{fig:rexmit_injected_loss}, causing the application-layer queue to overflow more.
This performance gap could be filled by parameter tuning.

We also consider an \emph{ideal} ``roofline'' protocol to calculate a fairly loose lower bound on the duty cycle.
This ideal protocol has the same header overhead as TCP, but learns which packets were lost for ``free,'' without using ACKs or running MMC.
Thus, it turns on its radio only to send out data and retransmit lost packets.
The real protocols have much higher duty cycles than the ideal protocol would have (Figure \ref{fig:rdc_injected_loss}), suggesting that a significant amount of their overhead stems from determining which packets were lost---polling the parent node for downstream TCP ACKs/CoAP responses.
This gap could be reduced by improving OpenThread's MMC protocol.
For example, rather than using a fixed sleep interval of 100 ms when an ACK is expected, one could use exponential backoff to increase the sleep interval if an ACK is not quickly received.
We leave exploring such ideas to future work.

\subsubsection{Performance in Lossy Conditions}\label{sec:lossy}

We compare the protocols over the course of a full day in our testbed, to study the impact of real wireless interference associated with human activity in an office.
We focus on \sys{} and CoAP since they were the most promising protocols from the previous experiment. To ensure that \sys{} and CoAP are subject to similar interference patterns, we (1) run them simultaneously, and (2) hardcode adjacent \sys{} and CoAP nodes to have the same first hop in the multihop topology.

\begin{figure}
    \centering
    \includegraphics[width=\linewidth]{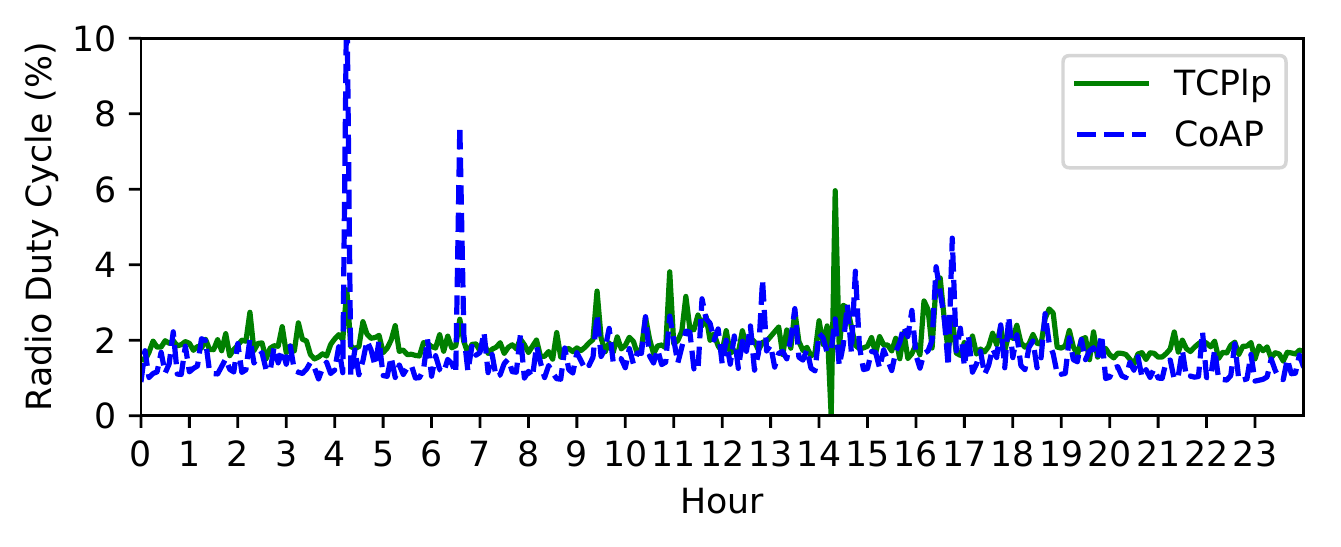}
    \caption{Radio duty cycle of TCP and CoAP in a lossy wireless environment, in one representative trial (losses are caused by natural human activity)}
    \label{fig:timeseries}
\end{figure}

\smallskip
\parhead{Improving Queue Management}
OpenThread's queue management interacts poorly with TCP in the presence of interference. When a duty-cycled leaf node sends a data request message to its parent, it turns its radio on and listens until it receives a reply (called an ``indirect message''). In OpenThread, the parent finishes sending its current frame (which may require link retries in the presence of interference), and then sends the indirect message. The duty-cycled leaf node keeps its radio on during this time, causing its radio duty cycle to increase. This is particularly bad for TCP, as its sliding window makes it more likely for the parent node to be in the middle of sending a frame when it receives a data request packet from a leaf node.
Thus, \textbf{we modified OpenThread to allow indirect messages to preempt the current frame \emph{in between link-layer retries}}, to minimize the time that duty-cycled leaf nodes must wait for a reply with their radios on. Both TCP and CoAP benefitted from this; TCP benefitted more because it suffered more from the problem to begin with.

\smallskip
\parhead{Power Consumption}
To improve power consumption for both TCP and CoAP, we adjusted parameters according to the lossy environment: (1) we enabled link-layer retries for indirect messages, (2) we decreased the data request timeout and performed link-layer retries more rapidly for indirect messages, to deliver them to leaves more quickly, and (3) given the high level of daytime interference, we decreased the MSS from five frames to three frames (as in \secref{sec:application}).

Figure \ref{fig:timeseries} depicts the radio duty cycle of TCP and CoAP for a trial representative of our overall results. \textbf{CoAP maintains a lower duty cycle than \sys{} outside of working hours, when there is less interference; \sys{} has a slightly lower duty cycle than CoAP during working hours, when there is more wireless interference.} \sys{}'s better performance at a higher loss rate is consistent with our results from \secref{sec:injected}. At a lower packet loss rate, TCP performs slightly worse than CoAP. This could be due to hidden terminal losses; more retries, on average, are required for indirect messages for TCP, causing leaf nodes to stay awake longer.
Overall, CoAP and \sys{} perform similarly (Table \ref{table:allday_performance}).

\begin{table}[t]
    \centering
    \begin{tabular}{| l || c | c | c |}\hline
         Protocol & Reliability & Radio DC & CPU DC \\\hline\hline
         \sys{} & 99.3\% & 2.29\% & 0.973\% \\\hline
         CoAP & 99.5\% & 1.84\% & 0.834\% \\\hline\hline
         Unrel., no batch & 93.4\% & 1.13\% & 0.52\% \\\hline
         Unrel., with batch & 95.3\% & 0.734\% & 0.30\% \\\hline
    \end{tabular}
    \caption{Performance in the testbed over a full day, averaged over multiple trials. The ideal protocol (\secref{sec:injected}) would have a radio DC of $\approx$ 0.63\%--0.70\% under similarly lossy conditions.}
    \label{table:allday_performance}
\end{table}

\subsubsection{Unreliable UDP}\label{s:reliability}

As a point of comparison, we repeat the sense-and-send experiment using a UDP-based protocol that \emph{does not provide reliability}. Concretely, we run CoAP in ``nonconfirmable'' mode, in which it does not use transport-layer ACKs or retransmissions.
The result is in the last two rows of Table \ref{table:allday_performance}. Compared to unreliable UDP, reliable approaches increase the radio/CPU duty cycle by 3x, in exchange for nearly
100\% reliability. That said, the corresponding decrease in battery life will be \emph{less} than 3x, because other sources of power consumption (reading from sensors, idle current) are also significant.

For other sense-and-send applications that sample at a lower rate, TCP and CoAP would see higher reliability (less application queue loss), but UDP would not similarly benefit (no application queue). Furthermore, the power consumption of TCP, CoAP, and unreliable UDP would all be closer together, given that the radio and CPU spend more time idle.

\subsection{Event Detection Application Scenario}\label{sec:evdetect}

Finally, we consider an application scenario where multiple flows compete for available bandwidth in an LLN. One such scenario is event detection: sensors wait until an interesting event occurs, at which point they report data upstream at a high data rate. Because such events tend to be correlated, multiple sensors send data simultaneously.

Nodes 12-15 in our testbed simultaneously transmit data to the EC2 instance (Figure \ref{fig:topology}), which measures the goodput of each flow. We use the same duty-cycling policy as in \secref{sec:senseandsend}. We divide each flow into 40-second intervals, measure the goodput in each interval, and compute the median and quartiles of goodput across all flows and intervals. The median gives a sense of aggregate goodput, and the quartiles gives a sense of fairness (quartiles close to the median are better).

\begin{figure}[t]
    \centering
    \hspace{-5ex}
    \begin{subfigure}[p]{0.37\linewidth}
        \includegraphics[width=\linewidth]{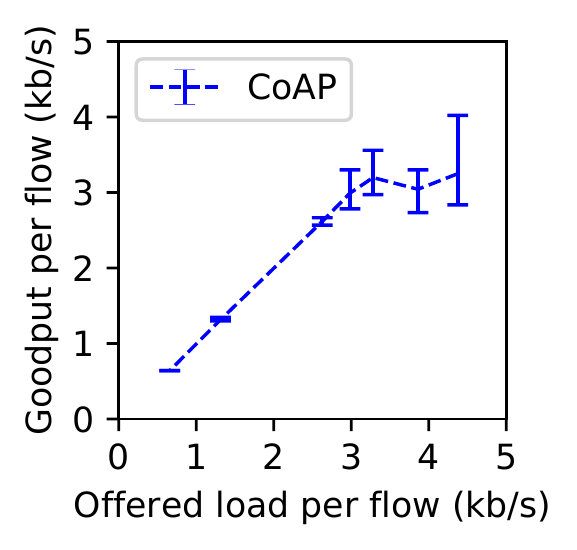}
        \label{fig:evdetect_coap}
    \end{subfigure}
    \hspace{-3ex}
    \begin{subfigure}[p]{0.37\linewidth}
        \includegraphics[width=\linewidth]{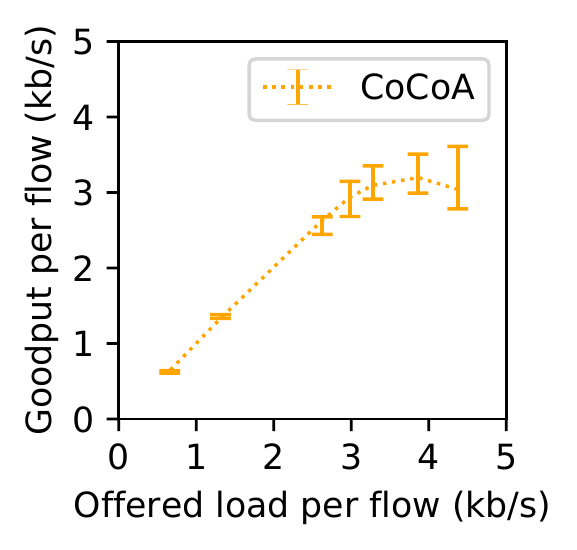}
        \label{fig:evdetect_cocoa}
    \end{subfigure}
    \hspace{-3ex}
    \begin{subfigure}[p]{0.37\linewidth}
        \includegraphics[width=\linewidth]{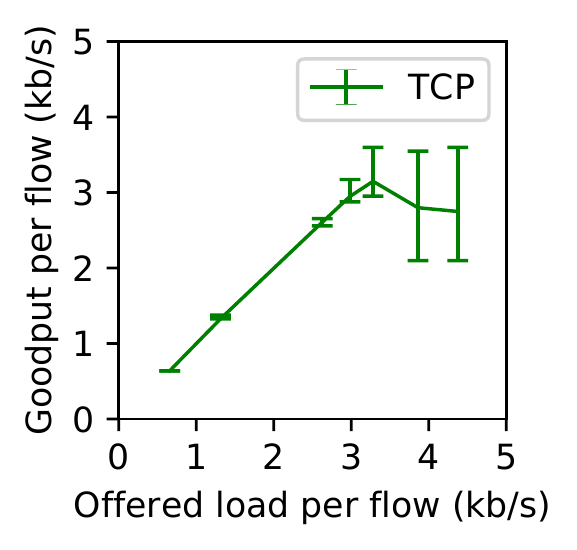}
        \label{fig:evdetect_tcp}
    \end{subfigure}
    \hspace{-5ex}

    \vspace{-12pt}
    \caption{CoAP, CoCoA, and TCP with four competing flows}
    \label{fig:evdetect}
\end{figure}

Figure \ref{fig:evdetect} shows the median and quartiles (as error bars) as the offered load increases. For small offered load, the per-flow goodput increases linearly. Once the aggregate load saturates the network, goodput declines slightly and the interquartile range increases, due to inefficiences in independent flows competing for bandwidth. \textbf{Overall, TCP performs similarly to CoAP and CoCoA, indicating that TCP's congestion control remains effective despite our observations in \secref{ssec:congestion} that it behaves differently in LLNs.}

\section{Conclusion}\label{sec:conclusion}

TCP is the \textit{de facto} reliability protocol in the Internet. Over the past 40 years, new physical-, datalink-, and application-layer protocols have evolved alongside TCP, and supporting good TCP performance was a consideration in their design. TCP is the obvious performance baseline for new transport-layer proposals. To warrant adoption, novel transports must be \emph{much} better than TCP in the intended application domain.

In contrast, when LLN research flourished two decades ago, LLN hardware could not run full-scale TCP. The original system architecture for networked sensors~\cite{hill2000system}, for example, targeted an 8-bit MCU with only 512 \emph{bytes} of memory. It naturally became taken for granted that TCP is too heavy for LLNs. Furthermore, contemporary research on TCP in WLANs~\cite{balakrishnan1995improving} suggested that TCP would perform poorly in LLNs even if the resource constraints were surmounted.

In revisiting the TCP question, after the resource constraints relaxed, we find that the expected pitfalls of wireless TCP actually do not carry over to LLNs. Although na\"ive TCP indeed performs poorly in LLNs, this is not due to fundamental problems with TCP as were observed in WLANs. Rather, it is caused by incompatibilities with a low-power link layer, which likely arose because canonical LLN protocols were developed in the absence of TCP considerations. We show how to fix these incompatibilities while preserving seamless interoperability with other TCP/IP networks.
This enables a viable TCP-based transport architecture for LLNs.

Our results have several implications for LLNs moving forward.
First, \textbf{the use of lightweight protocols that emulate part of TCP's functionality, like CoAP, needs to be reconsidered.} Protocol stacks like OpenThread should support full-scale TCP as an option. TCP should also serve as a benchmark to assess new LLN transport proposals.

Second, \textbf{full-scale TCP will influence the design of networked systems using LLNs.} Such systems are presently designed with application-layer gateways in mind (\secref{sec:tcp}). Using TCP/IP in the LLN itself would allow the use of commodity network management tools, like firewalls and NIDS. TCP would also allow the application-layer gateway to be replaced with a network-layer router, allowing clients to interact with LLN applications in much the same way as a Wi-Fi router allows users to interact with web applications. This is much more flexible than the status quo, where each LLN application needs application-specific functionality to be installed at the gateway~\cite{zachariah2015internet}. In cases where a new LLN transport protocol is truly necessary, the new protocol may be wise to consider the byte-stream \emph{abstraction} of TCP. This would allow the application-layer gateway to be replaced by a \emph{transport-layer gateway}. The mere presence of a transport layer, distinct from the application layer, goes a long way to providing interoperability with the rest of the Internet.

Third, \textbf{UDP-based protocols will still have a place in LLNs, just as they have a place in the Internet.} UDP is used for applications that benefit from greater control of segment transmission and loss response than TCP provides. These are typically real-time or multimedia applications where losing information is preferable to late delivery. It is entirely seemly for some sensing applications in LLNs, particularly those with similar real-time constraints, to transfer data using UDP-based protocols, even if TCP is an option. But TCP \emph{still} benefits such applications by providing a reliable channel for control information. For example, TCP may be used for device configuration, or to provide a shell for debugging, without yet another reliability protocol.

In summary, LLN-class devices are ready to become first-class citizens of the Internet. To this end, we believe that TCP should have a place in the LLN architecture moving forward, and that it will help put the ``I'' in IoT for LLN-class devices.

\section*{Acknowledgments}

We thank the anonymous reviewers, including in prior submissions, and our shepherd, Keith Winstein, for their invaluable feedback.
We are thankful to researchers in the BETS research group, including Kaifei Chen, Gabe Fierro, and Jack Kolb, for their feedback on early drafts of this paper, and to Prabal Dutta and Sylvia Ratnasamy for their advice and discussion. We also thank Albert Goto for his help with the LLN testbed.

This research is supported by the Department of Energy Grant DE-EE0007685, California Energy Commission, Intel Corporation, NSF Grant CPS-1239552, Fulbright Scholarship Program, UC Berkeley, and NSF Graduate Research Fellowship Program under Grant DGE-1752814.
Any opinions, findings, and conclusions or recommendations expressed in this material are those of the authors and do not necessarily reflect the views of the National Science Foundation.



\bibliographystyle{plain}
\bibliography{tcplp}

\begin{thebibliography}{100}

\bibitem{armdevicemanagementconnect}
Device management connect.
\newblock
  \url{https://www.arm.com/products/iot/pelion-iot-platform/device-management/connect}.
\newblock Accessed: 2018-09-09.

\bibitem{armjava}
Java speaks {CoAP}.
\newblock \url{https://community.arm.com/iot/b/blog/posts/java-speaks-coap}.
\newblock Accessed: 2018-09-09.

\bibitem{eclipse2014coap}
{MQTT} and {CoAP}, {IoT} protocols.
\newblock
  \url{https://www.eclipse.org/community/eclipse_newsletter/2014/february/article2.php}.
\newblock Accessed: 2018-09-09.

\bibitem{openthreadwebsite}
{OpenThread}.
\newblock \url{https://openthread.io/}.
\newblock Accessed: 2018-09-09.

\bibitem{cisco2018software}
Software configuration guide, {Cisco IOS} release 15.2(5)ex (catalyst digital
  building series switches).
\newblock
  \url{https://www.cisco.com/c/en/us/td/docs/switches/lan/catalyst_digital_building_series_switches/software/15-2_5_ex/configuration_guide/b_1525ex_consolidated_cdb_cg/b_1525ex_consolidated_cdb_cg_chapter_0111101.html}.
\newblock Accessed: 2018-09-09.

\bibitem{threadmembers}
Thread group.
\newblock \url{https://www.threadgroup.org/thread-group##OurMembers}.
\newblock Accessed: 2018-09-11.

\bibitem{threadproducts}
What is {Thread}.
\newblock \url{https://www.threadgroup.org/What-is-Thread##threadready}.
\newblock Accessed: 2018-09-12.

\bibitem{zeromq}
{ZeroMQ}.
\newblock \url{http://zeromq.org/}.
\newblock Accessed: 2019-01-29.

\bibitem{afanasyev2010host}
A.~Afanasyev, N.~Tilley, P.~Reiher, and L.~Kleinrock.
\newblock Host-to-host congestion control for {TCP}.
\newblock {\em IEEE Communications Surveys \& Tutorials}, 12(3), 2010.

\bibitem{alam2009crrt}
M.~M. Alam and C.~S. Hong.
\newblock {CRRT}: congestion-aware and rate-controlled reliable transport in
  wireless sensor networks.
\newblock {\em IEICE Transactions on Communications}, 92(1), 2009.

\bibitem{alizadeh2010data}
M.~Alizadeh, A.~Greenberg, D.~A. Maltz, J.~Padhye, P.~Patel, B.~Prabhakar,
  S.~Sengupta, and M.~Sridharan.
\newblock Data center {TCP} ({DCTCP}).
\newblock In {\em SIGCOMM}. ACM, 2010.

\bibitem{allman1999tcp}
M.~Allman.
\newblock {TCP} byte counting refinements.
\newblock {\em ACM SIGCOMM Computer Communication Review}, 29(3), 1999.

\bibitem{allman2003tcp}
M.~Allman.
\newblock {TCP} congestion control with appropriate byte counting ({ABC}).
\newblock {RFC} 3465, 2003.

\bibitem{allman2000enhancing}
M.~Allman, H.~Balakrishnan, and S.~Floyd.
\newblock Enhancing {TCP}'s loss recovery using limited transmit.
\newblock {RFC} 3042, 2000.

\bibitem{allman1999enhancing}
M.~Allman, D.~Glover, and L.~Sanchez.
\newblock Enhancing {TCP} over satellite channels using standard mechanisms.
\newblock {RFC} 2488, 1999.

\bibitem{allman1999estimating}
M.~Allman and V.~Paxson.
\newblock On estimating end-to-end network path properties.
\newblock {\em ACM SIGCOMM Computer Communication Review}, 29(4), 1999.

\bibitem{allman2009tcp}
M.~Allman, V.~Paxson, and E.~Blanton.
\newblock {TCP} congestion control.
\newblock {RFC} 5681, 2009.

\bibitem{andersen2016system}
M.~P Andersen, G.~Fierro, and D.~E. Culler.
\newblock System design for a synergistic, low power mote/{BLE} embedded
  platform.
\newblock In {\em IPSN}. ACM/IEEE, 2016.

\bibitem{arens2020measuring}
E.~Arens, A.~Ghahramani, R.~Przybyla, M.~P Andersen, S.~Min, T.~Peffer,
  P.~Raftery, M.~Zhu, V.~Luu, and H.~Zhang.
\newblock Measuring {3D} indoor air velocity via an inexpensive low-power
  ultrasonic anemometer.
\newblock {\em Energy and Buildings}, 211, 2020.

\bibitem{at86rf233}
Atmel Corporation.
\newblock {\em Low Power, {2.4GHz} Transceiver for {ZigBee}, {RF4CE}, {IEEE}
  802.15.4, {6LoWPAN}, and {ISM} Applications}, 2014.
\newblock Preliminary Datasheet.

\bibitem{ayadi2011tcp}
A.~Ayadi, P.~Maill{\'e}, and D.~Ros.
\newblock {TCP} over low-power and lossy networks: tuning the segment size to
  minimize energy consumption.
\newblock In {\em NTMS}. IEEE, 2011.

\bibitem{ayadi2011implementation}
A.~Ayadi, P.~Maill{\'e}, D.~Ros, L.~Toutain, and T.~Zheng.
\newblock Implementation and evaluation of a {TCP} header compression for
  {6LoWPAN}.
\newblock In {\em IWCMC}. IEEE, 2011.

\bibitem{ayadi2010tcp}
A.~Ayadi, D.~Ros, and L.~Toutain.
\newblock {TCP} header compression for {6LoWPAN}:
  draft-aayadi-6lowpan-tcphc-01.
\newblock Technical report, 2010.
\newblock \url{https://tools.ietf.org/id/draft-aayadi-6lowpan-tcphc-01}.

\bibitem{baccelli2018riot}
E.~Baccelli, C.~G{\"u}ndo{\u{g}}an, O.~Hahm, P.~Kietzmann, M.~S. Lenders,
  H.~Petersen, K.~Schleiser, T.~C. Schmidt, and M.~W{\"a}hlisch.
\newblock {RIOT}: an open source operating system for low-end embedded devices
  in the {IoT}.
\newblock {\em IEEE Internet of Things Journal}, 2018.

\bibitem{balakrishnan1997effects}
H.~Balakrishnan, V.~N. Padmanabhan, and R.~H. Katz.
\newblock The effects of asymmetry on {TCP} performance.
\newblock In {\em MobiCom}. ACM, 1997.

\bibitem{balakrishnan1997comparison}
H.~Balakrishnan, V.~N. Padmanabhan, S.~Seshan, and R.~H. Katz.
\newblock A comparison of mechanisms for improving {TCP} performance over
  wireless links.
\newblock {\em IEEE/ACM Transactions on Networking}, 5(6), 1997.

\bibitem{balakrishnan1995improving}
H.~Balakrishnan, S.~Seshan, E.~Amir, and R.~H. Katz.
\newblock Improving {TCP/IP} performance over wireless networks.
\newblock In {\em MobiCom}. ACM, 1995.

\bibitem{bershad1989lrpc}
B.~Bershad, T.~Anderson, E.~Lazowska, and H.~Levy.
\newblock Lightweight remote procedure call.
\newblock In {\em SOSP}. ACM, 1989.

\bibitem{betzler2016coap}
A.~Betzler, C.~Gomez, I.~Demirkol, and J.~Paradells.
\newblock {CoAP} congestion control for the {Internet of Things}.
\newblock {\em IEEE Communications Magazine}, 54(7), 2016.

\bibitem{borman2014tcp}
D.~Borman, B.~Braden, and V.~Jacobson.
\newblock {TCP} extensions for high performance.
\newblock (7323), 2014.

\bibitem{bormann2012coap}
C.~Bormann, A.~P. Castellani, and Z.~Shelby.
\newblock {CoAP}: An application protocol for billions of tiny internet nodes.
\newblock {\em IEEE Internet Computing}, 16(2), 2012.

\bibitem{borriello2000embedded}
G.~Borriello and R.~Want.
\newblock Embedded computation meets the world wide web.
\newblock {\em Communications of the ACM}, 43(5), 2000.

\bibitem{winter2012rpl}
A.~Brandt, J.~W. Hui, R.~Kelsey, P.~Levis, K.~Pister, R.~Struik, J.-P. Vasseur,
  and R.~Alexander.
\newblock {RPL}: {IPv6} routing protocol for low-power and lossy networks.
\newblock {RFC} 6550, 2012.

\bibitem{buettner2006x}
M.~Buettner, G.~V. Yee, E.~Anderson, and R.~Han.
\newblock {X-MAC}: a short preamble {MAC} protocol for duty-cycled wireless
  sensor networks.
\newblock In {\em SenSys}. ACM, 2006.

\bibitem{castellani2011web}
A.~P. Castellani, M.~Gheda, N.~Bui, M.~Rossi, and M.~Zorzi.
\newblock Web services for the {Internet of Things} through {CoAP} and {EXI}.
\newblock In {\em ICC}. IEEE, 2011.

\bibitem{clark1985structuring}
D.~D. Clark.
\newblock The structuring of systems using upcalls.
\newblock In {\em SOSP}. ACM, 1985.

\bibitem{clark1989analysis}
D.~D. Clark, V.~Jacobson, J.~Romkey, and H.~Salwen.
\newblock An analysis of {TCP} processing overhead.
\newblock {\em IEEE Communications magazine}, 27(6), 1989.

\bibitem{colitti2011evaluation}
W.~Colitti, K.~Steenhaut, N.~De Caro, B.~Buta, and V.~Dobrota.
\newblock Evaluation of constrained application protocol for wireless sensor
  networks.
\newblock In {\em LANMAN}. IEEE, 2011.

\bibitem{mqtt}
{MQTT} Community.
\newblock {MQTT}.
\newblock \url{http://mqtt.org}.
\newblock Accessed: January 25, 2018.

\bibitem{druschel1993fbufs}
P.~Druschel and L.~L. Peterson.
\newblock Fbufs: A high-bandwidth cross-domain transfer facility.
\newblock In {\em SOSP}. ACM, 1993.

\bibitem{cisco2013beyond}
P.~Duffy.
\newblock Beyond {MQTT}: A {Cisco} view on {IoT} protocols.
\newblock
  \url{https://blogs.cisco.com/digital/beyond-mqtt-a-cisco-view-on-iot-protocols}.
\newblock Accessed: 2018-09-09.

\bibitem{dunkels2003full}
A.~Dunkels.
\newblock Full {TCP/IP} for 8-bit architectures.
\newblock In {\em MobiSys}. ACM, 2003.

\bibitem{dunkels2003making}
A.~Dunkels, J.~Alonso, and T.~Voigt.
\newblock Making {TCP/IP} viable for wireless sensor networks.
\newblock {\em SICS Research Report}, 2003.

\bibitem{dunkels2004connecting}
A.~Dunkels, J.~Alonso, T.~Voigt, H.~Ritter, and J.~Schiller.
\newblock Connecting wireless sensornets with {TCP/IP} networks.
\newblock In {\em International Conference on Wired/Wireless Internet
  Communications}. Springer, 2004.

\bibitem{dunkels2004contiki}
A.~Dunkels, B.~Gr\"onvall, and T.~Voigt.
\newblock Contiki - a lightweight and flexible operating system for tiny
  networked sensors.
\newblock In {\em LCN}. IEEE, 2004.

\bibitem{duquennoy2015orchestra}
S.~Duquennoy, B.~Al~Nahas, O.~Landsiedel, and T.~Watteyne.
\newblock Orchestra: Robust mesh networks through autonomously scheduled
  {TSCH}.
\newblock In {\em SenSys}. ACM, 2015.

\bibitem{duquennoy2011lossy}
S.~Duquennoy, F.~{\"O}sterlind, and A.~Dunkels.
\newblock Lossy links, low power, high throughput.
\newblock In {\em SenSys}. ACM, 2011.

\bibitem{durvy2008making}
M.~Durvy, J.~Abeill\'{e}, P.~Wetterwald, C.~O’Flynn, B.~Leverett, E.~Gnoske,
  M.~Vidales, G.~Mulligan, N.~Tsiftes, N.~Finne, and A.~Dunkels.
\newblock Making sensor networks {IPv6} ready.
\newblock In {\em SenSys}. ACM, 2008.

\bibitem{dutta2010design}
P.~Dutta, S.~Dawson-Haggerty, Y.~Chen, C.-J.~M. Liang, and A.~Terzis.
\newblock Design and evaluation of a versatile and efficient receiver-initiated
  link layer for low-power wireless.
\newblock In {\em SenSys}. ACM, 2010.

\bibitem{estrin1999next}
D.~Estrin, R.~Govindan, J.~Heidemann, and S.~Kumar.
\newblock Next century challenges: Scalable coordination in sensor networks.
\newblock In {\em MobiCom}. ACM, 1999.

\bibitem{fall1996simulation}
K.~Fall and S.~Floyd.
\newblock Simulation-based comparisons of tahoe, reno and {SACK TCP}.
\newblock {\em ACM SIGCOMM Computer Communication Review}, 26(3), 1996.

\bibitem{floyd1994tcp}
S.~Floyd.
\newblock {TCP} and explicit congestion notification.
\newblock {\em ACM SIGCOMM Computer Communication Review}, 24(5), 1994.

\bibitem{floyd2003highspeed}
S.~Floyd.
\newblock {HighSpeed TCP} for large congestion windows.
\newblock {RFC} 3649, 2003.

\bibitem{floyd1993random}
S.~Floyd and V.~Jacobson.
\newblock Random early detection gateways for congestion avoidance.
\newblock {\em IEEE/ACM Transactions on Networking}, 1(4), 1993.

\bibitem{fouladi2018salsify}
S.~Fouladi, J.~Emmons, E.~Orbay, C.~Wu, R.~S. Wahby, and K.~Winstein.
\newblock Salsify: Low-latency network video through tighter integration
  between a video codec and a transport protocol.
\newblock In {\em NSDI}. USENIX, 2018.

\bibitem{freebsd}
The~FreeBSD Foundation.
\newblock {FreeBSD} 10.3, 2016.
\newblock \url{https://www.freebsd.org/releases/10.3R/announce.html}.

\bibitem{furst2016leveraging}
J.~Fürst, K.~Chen, M.~Aljarrah, and P.~Bonnet.
\newblock Leveraging physical locality to integrate smart appliances in
  non-residential buildings with ultrasound and bluetooth low energy.
\newblock In {\em IoTDI}. IEEE, 2016.

\bibitem{gerla1999tcp}
M.~Gerla, K.~Tang, and R.~Bagrodia.
\newblock {TCP} performance in wireless multi-hop networks.
\newblock In {\em WMCSA}. IEEE, 1999.

\bibitem{gnawali2009collection}
O.~Gnawali, R.~Fonseca, K.~Jamieson, D.~Moss, and P.~Levis.
\newblock Collection tree protocol.
\newblock In {\em SenSys}. ACM, 2009.

\bibitem{gomez2018tcp}
C.~Gomez, A.~Arcia-Moret, and J.~Crowcroft.
\newblock {TCP} in the {Internet of Things}: From ostracism to prominence.
\newblock {\em IEEE Internet Computing}, 22(1), 2018.

\bibitem{gont2011implementation}
F.~Gont and A.~Yourtchenko.
\newblock On the implementation of the {TCP} urgent mechanism.
\newblock {RFC} 6093, 2011.

\bibitem{grieco2004performance}
L.~A. Grieco and S.~Mascolo.
\newblock Performance evaluation and comparison of {Westwood+}, {New Reno}, and
  {Vegas} {TCP} congestion control.
\newblock {\em ACM SIGCOMM Computer Communication Review}, 34(2), 2004.

\bibitem{blemesh}
Bluetooth Mesh~Working Group.
\newblock Mesh profile v1.0, 2017.

\bibitem{thread}
Thread Group.
\newblock Thread, 2016.
\newblock \url{https://threadgroup.org}.

\bibitem{ha2008cubic}
S.~Ha, I.~Rhee, and L.~Xu.
\newblock {CUBIC}: A new {TCP}-friendly high-speed {TCP} variant.
\newblock {\em ACM SIGOPS Operating Systems Review}, 42(5), 2008.

\bibitem{henderson2012newreno}
T.~Henderson, S.~Floyd, A.~Gurtov, and Y.~Nishida.
\newblock The {NewReno} modification to {TCP}'s fast recovery algorithm.
\newblock {RFC} 6582, 2012.

\bibitem{hewage2015enabling}
K.~Hewage, S.~Duquennoy, V.~Iyer, and T.~Voigt.
\newblock Enabling {TCP} in mobile cyber-physical systems.
\newblock In {\em MASS}. IEEE, 2015.

\bibitem{hill2000system}
J.~Hill, R.~Szewczyk, A.~Woo, S.~Hollar, D.~E. Culler, and K.~Pister.
\newblock System architecture directions for networked sensors.
\newblock In {\em ASPLOS}. ACM, 2000.

\bibitem{jonathanhui}
J.~W. Hui.
\newblock Personal Communication.

\bibitem{hui2008ip}
J.~W. Hui and D.~E. Culler.
\newblock {IP} is dead, long live {IP} for wireless sensor networks.
\newblock In {\em SenSys}. ACM, 2008.

\bibitem{hull2004mitigating}
B.~Hull, K.~Jamieson, and H.~Balakrishnan.
\newblock Mitigating congestion in wireless sensor networks.
\newblock In {\em SenSys}. ACM, 2004.

\bibitem{im2015tcp}
H.~Im.
\newblock {\em {TCP} Performance Enhancement in Wireless Networks}.
\newblock PhD thesis, Seoul National University, 2015.

\bibitem{intanagonwiwat2000directed}
C.~Intanagonwiwat, R.~Govindan, and D.~Estrin.
\newblock Directed diffusion: A scalable and robust communication paradigm for
  sensor networks.
\newblock In {\em MobiCom}. ACM, 2000.

\bibitem{italianocalloutng}
D.~Italiano and A.~Motin.
\newblock Calloutng: a new infrastructure for timer facilities in the {FreeBSD}
  kernel.
\newblock In {\em AsiaBSDCon}, 2013.

\bibitem{iyer2005stcp}
Y.~G. Iyer, S.~Gandham, and S.~Venkatesan.
\newblock {STCP}: a generic transport layer protocol for wireless sensor
  networks.
\newblock In {\em ICCCN}. IEEE, 2005.

\bibitem{jacobson1988congestion}
V.~Jacobson.
\newblock Congestion avoidance and control.
\newblock In {\em SIGCOMM}. ACM, 1988.

\bibitem{jacobson1990compressing}
V.~Jacobson.
\newblock Compressing {TCP/IP} headers for low-speed serial links.
\newblock {RFC} 1144, 1990.

\bibitem{jin2004fast}
C.~Jin, D.~X. Wei, and S.~H. Low.
\newblock {FAST TCP}: motivation, architecture, algorithms, performance.
\newblock In {\em INFOCOM}. IEEE, 2004.

\bibitem{oma2016constrained}
S.~Johnson.
\newblock Constrained application protocol: {CoAP} is {IoT}'s `modern'
  protocol.
\newblock
  \url{https://www.omaspecworks.org/constrained-application-protocol-coap-is-iots-modern-protocol/},
  \url{https://internetofthingsagenda.techtarget.com/feature/Constrained-Application-Protocol-CoAP-is-IoTs-modern-protocol}.
\newblock Accessed: 2018-09-09.

\bibitem{ju2000efficient}
H.-T. Ju, M.-J. Choi, and J.~W. Hong.
\newblock An efficient and lightweight embedded web server for web-based
  network element management.
\newblock {\em International Journal of Network Management}, 10(5), 2000.

\bibitem{jung2017vibration}
D.~Jung, Z.~Zhang, and M.~Winslett.
\newblock Vibration analysis for {IoT} enabled predictive maintenance.
\newblock In {\em ICDE}. IEEE, 2017.

\bibitem{khalidi1995zerocopy}
Yousef~A. Khalidi and Moti~N. Thadani.
\newblock An efficient zero-copy {I/O} framework for {UNIX}.
\newblock Technical report, Mountain View, CA, USA, 1995.

\bibitem{kim2018system}
H.-S. Kim, M.~P Andersen, K.~Chen, S.~Kumar, W.~J. Zhao, K.~Ma, and D.~E.
  Culler.
\newblock System architecture directions for post-{SoC}/32-bit networked
  sensors.
\newblock In {\em SenSys}. ACM, 2018.

\bibitem{kim2017dt}
H.-S. Kim, H.~Cho, H.~Kim, and S.~Bahk.
\newblock {DT-RPL}: Diverse bidirectional traffic delivery through {RPL}
  routing protocol in low power and lossy networks.
\newblock {\em Computer Networks}, 126, 2017.

\bibitem{kim2015marketnet}
H.-S. Kim, H.~Cho, M.-S. Lee, J.~Paek, J.~Ko, and S.~Bahk.
\newblock {MarketNet}: An asymmetric transmission power-based wireless system
  for managing e-price tags in markets.
\newblock In {\em SenSys}. ACM, 2015.

\bibitem{kim2015measurement}
H.-S. Kim, H.~Im, M.-S. Lee, J.~Paek, and S.~Bahk.
\newblock A measurement study of {TCP} over {RPL} in low-power and lossy
  networks.
\newblock {\em Journal of Communications and Networks}, 17(6), 2015.

\bibitem{kim2019thread}
H.-S. {Kim}, S.~{Kumar}, and D.~E. {Culler}.
\newblock {Thread}/{OpenThread}: A compromise in low-power wireless multihop
  network architecture for the {Internet of Things}.
\newblock {\em IEEE Communications Magazine}, 57(7), 2019.

\bibitem{kim2007flush}
S.~Kim, R.~Fonseca, P.~Dutta, A.~Tavakoli, D.~E. Culler, P.~Levis, S.~Shenker,
  and I.~Stoica.
\newblock Flush: A reliable bulk transport protocol for multihop wireless
  networks.
\newblock In {\em SenSys}. ACM, 2007.

\bibitem{kim2007health}
S.~Kim, S.~Pakzad, D.~E. Culler, J.~Demmel, G.~Fenves, S.~Glaser, and M.~Turon.
\newblock Health monitoring of civil infrastructures using wireless sensor
  networks.
\newblock In {\em IPSN}. ACM/IEEE, 2007.

\bibitem{kovatsch2014californium}
M.~Kovatsch, M.~Lanter, and Z.~Shelby.
\newblock Californium: Scalable cloud services for the {Internet of Things}
  with {CoAP}.
\newblock In {\em IOT}. IEEE, 2014.

\bibitem{kumar2018bringing}
S.~Kumar, M.~P Andersen, H.-S. Kim, and D.~E. Culler.
\newblock Bringing full-scale {TCP} to low-power networks.
\newblock In {\em SenSys}. ACM, 2018.

\bibitem{kurose2013model}
J.~Kurose and K.~Ross.
\newblock {\em Computer Networking: A Top-Down Approach}, chapter~3, pages
  278--279.
\newblock 6th edition, 2013.

\bibitem{kushalnagar2007rfc}
N.~Kushalnagar, G.~Montenegro, and C.~Schumacher.
\newblock {IPv6} over low-power wireless personal area networks ({6LoWPANs}):
  Overview, assumptions, problem statement, and goals.
\newblock {RFC} 4919, 2007.

\bibitem{levis2003tossim}
P.~Levis, N.~Lee, M.~Welsh, and D.~E. Culler.
\newblock {TOSSIM}: Accurate and scalable simulation of entire {TinyOS}
  applications.
\newblock In {\em SenSys}. ACM, 2003.

\bibitem{levis2005tinyos}
P.~Levis, S.~Madden, J.~Polastre, R.~Szewczyk, K.~Whitehouse, A.~Woo, D.~Gay,
  J.~Hill, M.~Welsh, E.~Brewer, and D.~E. Culler.
\newblock {\em {TinyOS}: An operating system for sensor networks}.
\newblock 2005.

\bibitem{levis2004trickle}
P.~Levis, N.~Patel, D.~E. Culler, and S.~Shenker.
\newblock Trickle: A self-regulating algorithm for code propagation and
  maintenance in wireless sensor networks.
\newblock In {\em NSDI}. USENIX, 2004.

\bibitem{levy2016beetle}
A.~A. Levy, J.~Hong, L.~Riliskis, P.~Levis, and K.~Winstein.
\newblock Beetle: Flexible communication for bluetooth low energy.
\newblock In {\em MobiSys}. ACM, 2016.

\bibitem{li2005lyranet}
Y.-C. Li and M.-L. Chiang.
\newblock {LyraNET}: a zero-copy {TCP/IP} protocol stack for embedded operating
  systems.
\newblock In {\em RTCSA}. IEEE, 2005.

\bibitem{liang2010surviving}
C.-J.~M. Liang, N.~B. Priyantha, J.~Liu, and A.~Terzis.
\newblock Surviving {Wi-Fi} interference in low power {ZigBee} networks.
\newblock In {\em SenSys}. ACM, 2010.

\bibitem{ludwig2003tcp}
R.~Ludwig, A.~Gurtov, and F.~Khafizov.
\newblock {TCP} over second ({2.5G}) and third ({3G}) generation wireless
  networks.
\newblock {RFC} 3481, 2003.

\bibitem{maeda1993protocol}
C.~Maeda and B.~N. Bershad.
\newblock Protocol service decomposition for high-performance networking.
\newblock In {\em SOSP}. ACM, 1993.

\bibitem{mainwaring2002wireless}
A.~Mainwaring, D.~E. Culler, J.~Polastre, R.~Szewczyk, and J.~Anderson.
\newblock Wireless sensor networks for habitat monitoring.
\newblock In {\em WSNA}. ACM, 2002.

\bibitem{mathis1997macroscopic}
M.~Mathis, J.~Semke, J.~Mahdavi, and T.~Ott.
\newblock The macroscopic behavior of the {TCP} congestion avoidance algorithm.
\newblock {\em ACM SIGCOMM Computer Communication Review}, 27(3), 1997.

\bibitem{mcewen2013risking}
A.~McEwen.
\newblock Risking a compuserve of things.
\newblock
  \url{https://mcqn.com/posts/wuthering-bytes-slides-risking-a-compuserve-of-things/}.
\newblock Accessed: 2018-12-08.

\bibitem{montenegro2008lowpan}
G.~Montenegro, N.~Kushalnagar, J.~W. Hui, and D.~E. Culler.
\newblock Transmission of {IPv6} packets over {IEEE} 802.15.4 networks.
\newblock {RFC} 4944, 2007.

\bibitem{openthread}
Google Nest.
\newblock {OpenThread}, 2017.
\newblock \url{https://github.com/openthread/openthread}.

\bibitem{osterlind2008approaching}
F.~{\"O}sterlind and A.~Dunkels.
\newblock Approaching the maximum 802.15.4 multi-hop throughput.
\newblock In {\em HotEmNets}. ACM, 2008.

\bibitem{padhye1998modeling}
J.~Padhye, V.~Firoiu, D.~Towsley, and J.~Kurose.
\newblock Modeling {TCP} throughput: A simple model and its empirical
  validation.
\newblock {\em ACM SIGCOMM Computer Communication Review}, 28(4), 1998.

\bibitem{paek2007rcrt}
J.~Paek and R.~Govindan.
\newblock {RCRT}: Rate-controlled reliable transport for wireless sensor
  networks.
\newblock In {\em SenSys}. ACM, 2007.

\bibitem{pang2007reliable}
Q.~Pang, V.~W.~S. Wong, and V.~C.~M. Leung.
\newblock Reliable data transport and congestion control in wireless sensor
  networks.
\newblock {\em International Journal of Sensor Networks}, 3(1), 2008.

\bibitem{paxson1999known}
V.~Paxson, M.~Allman, S.~Dawson, W.~Fenner, J.~Griner, I.~Heavens, K.~Lahey,
  J.~Semke, and B.~Volz.
\newblock Known {TCP} implementation problems.
\newblock {RFC} 2525, 1999.

\bibitem{polastre2004versatile}
J.~Polastre, J.~Hill, and D.~E. Culler.
\newblock Versatile low power media access for wireless sensor networks.
\newblock In {\em SenSys}. ACM, 2004.

\bibitem{polastre2005telos}
J.~Polastre, R.~Szewczyk, and D.~E. Culler.
\newblock {Telos}: Enabling ultra-low power wireless research.
\newblock In {\em IPSN}. ACM/IEEE, 2005.

\bibitem{rahman2008wireless}
M.~A. Rahman, A.~El~Saddik, and W.~Gueaieb.
\newblock {\em Wireless Sensor Network Transport Layer: State of the Art}.
\newblock 2008.

\bibitem{raiciu2012mptcp}
C.~Raiciu, C.~Paasch, S.~Barre, A.~Ford, M.~Honda, F.~Duchene, O.~Bonaventure,
  and M.~Handley.
\newblock How hard can it be? {Designing} and implementing a deployable
  multipath {TCP}.
\newblock In {\em NSDI}. USENIX, 2012.

\bibitem{ramaiah2010improving}
A.~Ramaiah, M.~Stewart, and M.~Dalal.
\newblock Improving {TCP}'s robustness to blind in-window attacks.
\newblock {RFC} 5961, 2010.

\bibitem{rathnayaka2013wireless}
A.~J.~D. Rathnayaka and V.~M. Potdar.
\newblock Wireless sensor network transport protocol: A critical review.
\newblock {\em Journal of Network and Computer Applications}, 36(1), 2013.

\bibitem{sankarasubramaniam2003esrt}
Y.~Sankarasubramaniam, {\"O}.~B. Akan, and I.~F. Akyildiz.
\newblock Esrt: event-to-sink reliable transport in wireless sensor networks.
\newblock In {\em MobiHoc}. ACM, 2003.

\bibitem{santos2015personal}
D.~F.~S. Santos, H.~O. Almeida, and A.~Perkusich.
\newblock A personal connected health system for the {Internet of Things} based
  on the {Constrained Application Protocol}.
\newblock {\em Computers \& Electrical Engineering}, 44, 2015.

\bibitem{schmid2010disentangling}
T.~Schmid, R.~Shea, M.~B. Srivastava, and P.~Dutta.
\newblock Disentangling wireless sensing from mesh networking.
\newblock In {\em HotEmNets}, 2010.

\bibitem{seitz2017enabling}
K.~Seitz, S.~Serth, K.-F. Krentz, and C.~Meinel.
\newblock Enabling en-route filtering for end-to-end encrypted {CoAP} messages.
\newblock In {\em SenSys}. ACM, 2017.

\bibitem{semke1998automatic}
J.~Semke, J.~Mahdavi, and M.~Mathis.
\newblock Automatic {TCP} buffer tuning.
\newblock In {\em SIGCOMM}. ACM, 1998.

\bibitem{shelby2014constrained}
Z.~Shelby, K.~Hartke, and C.~Bormann.
\newblock The constrained application protocol ({CoAP}).
\newblock {RFC} 7252, 2014.

\bibitem{snoeren2000end}
A.~C. Snoeren and H.~Balakrishnan.
\newblock An end-to-end approach to host mobility.
\newblock In {\em MobiCom}. ACM, 2000.

\bibitem{stann2003rmst}
F.~Stann and J.~Heidemann.
\newblock {RMST}: Reliable data transport in sensor networks.
\newblock In {\em SNPA}. IEEE, 2003.

\bibitem{stathopoulos2005mote}
T.~Stathopoulos, L.~Girod, J.~Heidemann, and D.~Estrin.
\newblock Mote herding for tiered wireless sensor networks.
\newblock Technical Report~58, University of California, Los Angeles, Center
  for Embedded Networked Computing, December 2005.

\bibitem{szewczyk2004lessons}
R.~Szewczyk, J.~Polastre, A.~Mainwaring, and D.~E. Culler.
\newblock Lessons from a sensor network expedition.
\newblock In H.~Karl, A.~Wolisz, and A.~Willig, editors, {\em EWSN}. Springer
  Berlin Heidelberg, 2004.

\bibitem{vasseur2014rfc}
J.-P. Vasseur.
\newblock Terms used in routing for low-power and lossy networks.
\newblock {RFC} 7102, 2014.

\bibitem{villaverde2012constrained}
B.~C. Villaverde, D.~Pesch, R.~De~Paz Alberola, S.~Fedor, and M.~Boubekeur.
\newblock {Constrained Application Protocol} for low power embedded networks: A
  survey.
\newblock In {\em IMIS}. IEEE, 2012.

\bibitem{wan2002psfq}
C.-Y. Wan, A.~T. Campbell, and L.~Krishnamurthy.
\newblock {PSFQ}: a reliable transport protocol for wireless sensor networks.
\newblock In {\em WSNA}. ACM, 2002.

\bibitem{wan2003coda}
C.-Y. Wan, S.~B. Eisenman, and A.~T. Campbell.
\newblock {CODA}: Congestion detection and avoidance in sensor networks.
\newblock In {\em SenSys}. ACM, 2003.

\bibitem{wang2005issues}
C.~Wang, K.~Sohraby, Y.~Hu, B.~Li, and W.~Tang.
\newblock Issues of transport control protocols for wireless sensor networks.
\newblock In {\em ICCCAS}. IEEE, 2005.

\bibitem{windley2014compuserve}
P.~Windley.
\newblock The compuserve of things.
\newblock
  \url{http://www.windley.com/archives/2014/04/the_compuserve_of_things.shtml}.
\newblock Accessed: 2018-12-08.

\bibitem{winstein2012mosh}
K.~Winstein and H.~Balakrishnan.
\newblock Mosh: An interactive remote shell for mobile clients.
\newblock In {\em ATC}. USENIX, 2012.

\bibitem{winstein2013stochastic}
K.~Winstein, A.~Sivaraman, and H.~Balakrishnan.
\newblock Stochastic forecasts achieve high throughput and low delay over
  cellular networks.
\newblock In {\em NSDI}. USENIX, 2013.

\bibitem{woo2001transmission}
A.~Woo and D.~E. Culler.
\newblock A transmission control scheme for media access in sensor networks.
\newblock In {\em MobiCom}. ACM, 2001.

\bibitem{wright1995tcpvol2chap2}
G.~R. Wright and W.~R. Stevens.
\newblock {\em {TCP/IP} Illustrated}, volume~2, chapter~2.
\newblock 1995.

\bibitem{xu2004wisden}
N.~Xu, S.~Rangwala, K.~K. Chintalapudi, D.~Ganesan, A.~Broad, R.~Govindan, and
  D.~Estrin.
\newblock A wireless sensor network for structural monitoring.
\newblock In {\em SenSys}. ACM, 2004.

\bibitem{ye2002energy}
W.~Ye, J.~Heidemann, and D.~Estrin.
\newblock An energy-efficient {MAC} protocol for wireless sensor networks.
\newblock In {\em INFOCOM}. IEEE, 2002.

\bibitem{ye2004medium}
W.~Ye, J.~Heidemann, and D.~Estrin.
\newblock Medium access control with coordinated adaptive sleeping for wireless
  sensor networks.
\newblock {\em IEEE/ACM Transactions on Networking}, 12(3), 2004.

\bibitem{zachariah2015internet}
T.~Zachariah, N.~Klugman, B.~Campbell, J.~Adkins, N.~Jackson, and P.~Dutta.
\newblock The {Internet of Things} has a gateway problem.
\newblock In {\em HotMobile}. ACM, 2015.

\bibitem{zhang2005reliable}
H.~Zhang, A.~Arora, Y.-R. Choi, and M.~G. Gouda.
\newblock Reliable bursty convergecast in wireless sensor networks.
\newblock In {\em MobiHoc}. ACM, 2005.

\bibitem{zhang1986why}
L.~Zhang.
\newblock Why {TCP} timers don’t work well.
\newblock {\em ACM SIGCOMM Computer Communication Review}, 16(3), 1986.

\bibitem{zheng2011tcp}
T.~Zheng, A.~Ayadi, and X.~Jiang.
\newblock {TCP} over {6LoWPAN} for industrial applications: An experimental
  study.
\newblock In {\em NTMS}. IEEE, 2011.

\end{thebibliography}

\appendix
\section{Impact of Network Stack Design}\label{sec:netstack}

As mentioned in \secref{sec:implementation}, we made nontrivial modifications to FreeBSD's TCP stack to port it to each embedded operating system and embedded network stack. Below we provide additional information about these changes, and about our implementations for platforms other than Hamilton/OpenThread.

\smallskip
\subsection{Concurrency Model}\label{s:concurrency}

\parhead{GNRC and OpenThread (RIOT OS)}
RIOT OS provides threads as the basic unit of concurrency. Asynchronous interaction with hardware is done by interrupt handlers that preempt the current thread, perform a short operation in the interrupt context, and signal a related thread to perform any remaining operation outside of interrupt context. Then the thread is placed on the RIOT OS scheduler queue and is scheduled for execution depending on its priority.

The GNRC network stack for RIOT OS runs each network layer (or module) in a separate thread. Each thread has a priority and can be preempted by a thread with higher priority or by an interrupt. The thread for a lower network layer has higher priority than the thread for a higher layer.

The port of OpenThread for RIOT OS handles received packets in one thread and sends packets from another thread, where the thread for received packets has higher priority~\cite{kim2018system}. The rationale for this design is to ensure timely processing of received packets at the radio, which is especially important in the context of a high-throughput flow.

To adapt \sys{} for GNRC, we run the FreeBSD implementation as a single TCP-layer thread, whose priority is between that of the application-layer thread and the IPv6-layer thread. To adapt \sys{} for OpenThread on RIOT OS, we call the TCP protocol logic (\texttt{tcp\_input()}) at the appropriate point along the receive path, and send packets from the TCP protocol logic (\texttt{tcp\_output()}) using the established send path. As explained in \appref{s:timers}, we also use an additional thread for timer callbacks in RIOT OS.

Given that TCP state can be accessed concurrently from multiple threads---the TCP thread (GNRC) or receive thread (OpenThread), the application thread(s), and timer callbacks---we needed to synchronize access to it. The FreeBSD implementation allows fine-grained locking of connection state to allow different connections to be serviced in parallel on different CPUs. Given that low-power embedded sensors typically have only one CPU, however, we opted for simplicity, instead using a single global TCP lock for \sys{}.

\smallskip
\parhead{BLIP (TinyOS)}
TinyOS uses an event-driven concurrency model based on split-phase operations, consisting of an event loop that executes on a \emph{single} stack.
For concurrency, TinyOS provides three types of unique operations: \emph{commands} and \emph{events}, which are executed immediately, and \emph{tasks}, which are scheduled for execution after all preceding tasks are completed.
An interrupt handler may preempt the current function, perform a short operation in the interrupt context using \emph{asynchronous} events and commands, and \emph{post} a task to perform any remaining computation later.
To adapt the thread-based FreeBSD implementation to the event-driven TinyOS, we execute the primary functions of FreeBSD, such as \texttt{tcp\_output()} and \texttt{tcp\_input()}, within \emph{tasks} outside of interrupt context. Because tasks in TinyOS cannot preempt each other, we remove the locking present in the FreeBSD TCP implementation.

\subsection{Timer Event Management}\label{s:timers}

Given that many TCP operations are based on timer events, achieving correct timer operation is important. For example, if an RTO timer event is dropped by the embedded operating system, the RTO timer will not be rescheduled, and the connection may hang.

For a simple and stable operation, many existing embedded TCP stacks, including the uIP, lwIP, and BLIP TCP stacks, rely on a periodic, fixed-interval clock in order to check for expired timeouts.
Instead, \sys{} uses one-shot tickless timers as FreeBSD 10.3 does~\cite{italianocalloutng}, which is beneficial in two ways: (1) When there are no scheduled timers, the tickless timers allow the CPU to sleep, rather than being needlessly woken up at a fixed interval, resulting in lower energy consumption~\cite{kim2018system}. (2) Unlike fixed periodic timers, which can only be serviced on the next tick after they expire, tickless timers can be serviced as soon as they expire.
To obtain these advantages, however, an embedded operating system must robustly manage asynchronous timer callbacks.

TinyOS has a single event queue maintained by the scheduler. The semantics of TinyOS guarantee that a task can exist in the event queue only once, even if it is \emph{posted} (i.e., scheduled for execution) multiple times before executing. Therefore, the event queue can be sized appropriately at compile-time to not overflow. Furthermore, TinyOS handles received packets in a separate queue than tasks. This ensures that TCP timer callbacks will not be dropped.

This is not the case for RIOT OS. Timer callbacks either handle the timer entirely in interrupt context, or put an event on a thread's message queue, so that the thread performs the required callback operation. Each network protocol supported by RIOT OS has a single thread.
Because a thread's message queue in RIOT OS is used to hold both received packets and timer events, there is no guarantee when a timer expires that there is enough space in the thread message queue to accept a timer event; if there is not enough space, RIOT OS drops the timer event. Furthermore, if a timer expires multiple times before its event is handled by the thread, multiple events for the same timer can exist simultaneously in the queue; \emph{we cannot find an upper bound on the number of slots in the message queue used by a single timer}.
To provide robust TCP operation on RIOT OS, we create a second thread used exclusively for TCP timers. We handle timers similarly to TinyOS' \emph{post} operation, by preventing the message queue from having multiple callback events of a single timer. This eliminates the possibility of timer event drops.

\begin{table}[t]
    \centering
    \setlength\tabcolsep{5pt}
    \begin{tabular}{| l || c | c | c |}\hline
         & Protocol & Event Sched. & User Library \\\hline\hline
         ROM & 21352 B & 1696 B & 5384 B \\\hline
         RAM (Active) & 488 B & 40 B & 36 B \\\hline
         RAM (Passive) & 16 B & 16 B & 36 B\\\hline
    \end{tabular}
    \caption{Memory usage of \sys{} on TinyOS. Our implementation of \sys{} spans three modules: (1) protocol implementation, (2) event scheduler that injects callbacks into userspace, and (3) userland library.}
    \label{table:memory_tinyos}
\end{table}

\subsection{Memory Usage: Connection State}
To complement Table \ref{table:memory_riot}, which shows \sys{}'s memory footprint on RIOT OS, we include Table \ref{table:memory_tinyos}, which shows \sys{}'s memory footprint on TinyOS.

\subsection{Performance Comparison}\label{sec:direct}

We consider TCP goodput between two embedded nodes over the IEEE 802.15.4 link, over a single hop without any border router, as we did in \secref{sec:upper_bound}. We are able to produce a 63 kb/s goodput over a TCP connection between two Hamilton motes using RIOT's GNRC network stack. For comparison, we are able to achieve 71 kb/s using the BLIP stack on Firestorm, and 75 kb/s using the OpenThread network stack with RIOT OS on Hamilton.
\textbf{This suggests that our results are reproducible across multiple platforms and embedded network stacks.} The minor performance degradation in GNRC is partially explained by its greater header overhead due to implementation differences, and by its IPC-based thread-per-layer concurrency architecture, which has known inefficiencies~\cite{clark1985structuring}. This suggests that the implementation of the underlying network stack, particularly with regard to concurrency, could affect TCP performance in LLNs.

\section{Comparison of Features in Embedded TCP Implementations}\label{sec:features}

Table \ref{table:feature_comparison} compares the featureset of \sys{} to features in embedded TCP stacks. The TCP implementations in uIP and BLIP lack features core to TCP. uIP allows only one unACKed in-flight segment, eschewing TCP's sliding window. BLIP does not implement RTT estimation or congestion control. The TCP implementation in GNRC lacks features such as TCP timestamps, selective ACKs, and delayed ACKs, which are present in most full-scale TCP implementations.

\begin{table}[t]
    \centering
    \begin{tabular}{| l || c | c | c || c |} \hline
    & uIP & BLIP & GNRC & \sys{} \\ \hline\hline
    Flow Control & Yes & Yes & Yes & Yes\\ \hline
    Congestion Control & N/A & No & Yes & Yes\\ \hline
    RTT Estimation & Yes & No & Yes & Yes\\ \hline
    MSS Option & Yes & No & Yes & Yes\\ \hline
    OOO Reassembly & No & No & Yes & Yes\\ \hline
    TCP Timestamps & No & No & No & Yes\\ \hline
    Selective ACKs & No & No & No & Yes\\ \hline
    Delayed ACKs & No & No & No & Yes\\ \hline
    \end{tabular}
    \caption{Comparison of core features among embedded TCP stacks: uIP (Contiki), BLIP (TinyOS), GNRC (RIOT), and \sys{} \emph{(this paper)}}
    \label{table:feature_comparison}
\end{table}

\smallskip
\parhead{Benefits of full-scale TCP}
In addition to supporting the protocol-level features summarized in Table \ref{table:feature_comparison}, \sys{} is likely more robust than other embedded TCP stacks because it is based on a well-tested TCP implementation.
While seemingly minor, some details, implemented incorrectly by TCP stacks, have had important consequences for TCP's behavior~\cite{paxson1999known}.
\sys{} benefits from a thorough implementation of each aspect of TCP.

For example, \sys{}, by virtue of using the FreeBSD TCP implementation, benefits from a robust implementation of congestion control.
\sys{} implements not only the basic New Reno algorithm, but also Explicit Congestion Notification~\cite{floyd1994tcp}, Appropriate Byte Counting~\cite{allman1999tcp, allman2003tcp} and Limited Transmissions~\cite{allman2000enhancing}.
It also inherits from FreeBSD heuristics to identify and correct ``bad retransmissions'' (as in \S{}2.8 of \cite{allman1999estimating}): if, after a retransmission, the corresponding ACK is received very soon (within $\frac{\rtt{}}{2}$ of the retransmission), the ACK is assumed to correspond to the originally transmitted segment as opposed to the retransmission.
The FreeBSD implementation and \sys{} recover from such ``bad retransmissions'' by restoring \cwnd{} and \ssthresh{} to their former values before the packet loss.
Aside from congestion control, \sys{} benefits from header prediction~\cite{clark1989analysis}, which introduces a ``fast code path'' to process common-case TCP segments (in-sequence data and ACKs) more efficiently, and Challenge ACKs~\cite{ramaiah2010improving}, which make it more difficult for an attacker to inject an RST into a TCP connection.

Enhancements such as these make us more confident that our observed results are fundamental to TCP, as opposed to artifacts of poor implementation. Furthermore, they allow us to focus on performance problems arising from the challenges of LLNs, as opposed to general TCP-related challenges that the research community has already solved in the context of traditional networks and operating systems.

\section{Derivation of TCP Model}\label{sec:derivation}

This appendix provides the derivation of Equation \ref{eq:lln_model}, the model of TCP performance proposed in \secref{sec:model}.

We think of a TCP flow as a sequence of bursts. A \emph{burst} is a sequence of full windows of data successfully transferred, which ends in a packet loss. After this loss, the flow spends some time recovering from the packet loss, which we call a \emph{rest}. Then, the next burst begins. Let $w$ be the size of TCP's flow window, measured in segments (for our experiments in \secref{ssec:congestion}, we would have $w=4$). Define $b$ as the average number of windows sent in a burst. The goodput of TCP is the number of bytes sent in each burst, which is $w \cdot b \cdot \mss$, divided by the duration of each burst. A burst lasts for the time to transmit $b$ windows of data, plus the time to recover from the packet loss that ended the burst. The time to transmit $b$ windows is $b \cdot \rtt$. We define $\trec$ to be the time to recover from the packet loss. Then we have
\begin{equation}
    B = \frac{w \cdot b \cdot \mss}{b \cdot \rtt + \trec}.
\end{equation}
The value of $b$ depends on the packet loss rate. We define a new variable, $\pwin$, which denotes the probability that at least one packet in a window is lost. Then $b = \frac{1}{\pwin}$.

To complete the model, we must estimate $\trec$ and $\pwin$.

The value of $\trec$ depends on whether the retransmission timer expires (called an RTO) or a fast retransmission is performed. If an RTO occurs, the total time lost is the excess time budgeted to the retransmit timer beyond one RTT, plus the time to retransmit the lost segments. We denote the time budgeted to the retransmit timer as ETO. So the total time lost due to a timeout, assuming it takes about 2 RTTs to recover lost segments, would be $(\eto - \rtt) + 2\cdot\rtt = \eto + \rtt$.  After a fast retransmission, TCP enters a ``fast recovery'' state~\cite{allman2009tcp, henderson2012newreno}. Fast recovery requires buffer space to be effective, however. In particular, if the buffer contains only four TCP segments, then the lost packet, and three packets afterward which resulted in duplicate ACKs, account for the entire send buffer; therefore, TCP cannot send new data during fast recovery, and instead stalls for one RTT, until the ACK for the fast retransmission is received. In contrast, choosing a larger send buffer will allow fast recovery to more effectively mask this loss~\cite{semke1998automatic}.

As discussed in \secref{ssec:congestion}, these two types of losses may be caused by different factors. Therefore, we do not attempt to distinguish them on basis of probability. Instead, we use a very simple model: $\trec = \ell \cdot \rtt$. The constant $\ell$ can be chosen to describe the number of ``productive'' RTTs lost due to a packet loss. Based on the estimates above, choosing $\ell = 2$ seems reasonable for our experiments in \secref{sec:multihop} which used a buffer size of four segments.

To model $\pwin$, we assume that, in each window, segment losses are independent. This gives us $\pwin = 1 - (1 - p)^w$, where $p$ is the probability of an individual segment being lost (after link retries). Because $p$ is likely to be small (less than 20\%), we apply the approximation that $(1 - x)^a \approx 1 - ax$ for small $x$. This gives us $\pwin \approx wp$.

Applying these equations for $\trec$ and $\pwin$, along with some minor algebraic manipulation to put our equation in a similar form to Equation \ref{eq:newreno_model}, we obtain our model for TCP performance in LLNs, for small $w$ and $p$:
\begin{equation}\label{eq:lln_full_model}
    B = \frac{\mss}{\rtt} \cdot \frac{1}{\frac{1}{w} + \ell p}
\end{equation}
Equation \ref{eq:lln_model}, stated in \secref{sec:model}, takes $\ell = 2$, as discussed above.

\smallskip
\parhead{Generalizing the model}
In \secref{sec:application}, we generally use a smaller MSS (3 frames) than we used in \secref{sec:multihop}. Furthermore, duty-cycling increases the RTT. It is natural to ask whether our conclusions in \secref{sec:multihop}, on which the model is based, still hold in this setting. With a sleep interval of 100 ms, we qualitatively observed that, although \cwnd{} tends to recover more slowly after loss, due to the smaller MSS and larger RTT, it is still ``maxed out'' past the BDP most of the time. Therefore, we expect our conclusion, that TCP is more resilient to packet loss, to also apply in this setting.

One may consider adapting our model for this setting by choosing a larger value of $\ell$ to reflect the fact that \cwnd{} recovers from loss less quickly due to the smaller MSS. It is possible, however, that one could derive a better model by explicitly modeling the phase when \cwnd{} is recovering, similar to other existing TCP models (in contrast to our model above, where we assume that the TCP flow is binary---either transmitting at a full window, or in backoff after loss). We leave exploration of this idea to future work.

\section{Anemometry: An LLN Application}\label{app:anemometer}

An \emph{anemometer} is a sensor that measures air velocity. Ane-mometers may be deployed in a building to diagnose problems with the Heating, Ventilation, and Cooling system (HVAC), and also to collect air flow measurements for improved HVAC control. This requires anemometers in difficult-to-reach locations, such as in air flow ducts, where it is infeasible to run wires. Therefore, anemometers must be battery-powered and must transmit readings wirelessly, making LLNs attractive.

\begin{figure}[t]
    \centering
    \begin{subfigure}[p]{0.25\linewidth}
        \includegraphics[width=\linewidth]{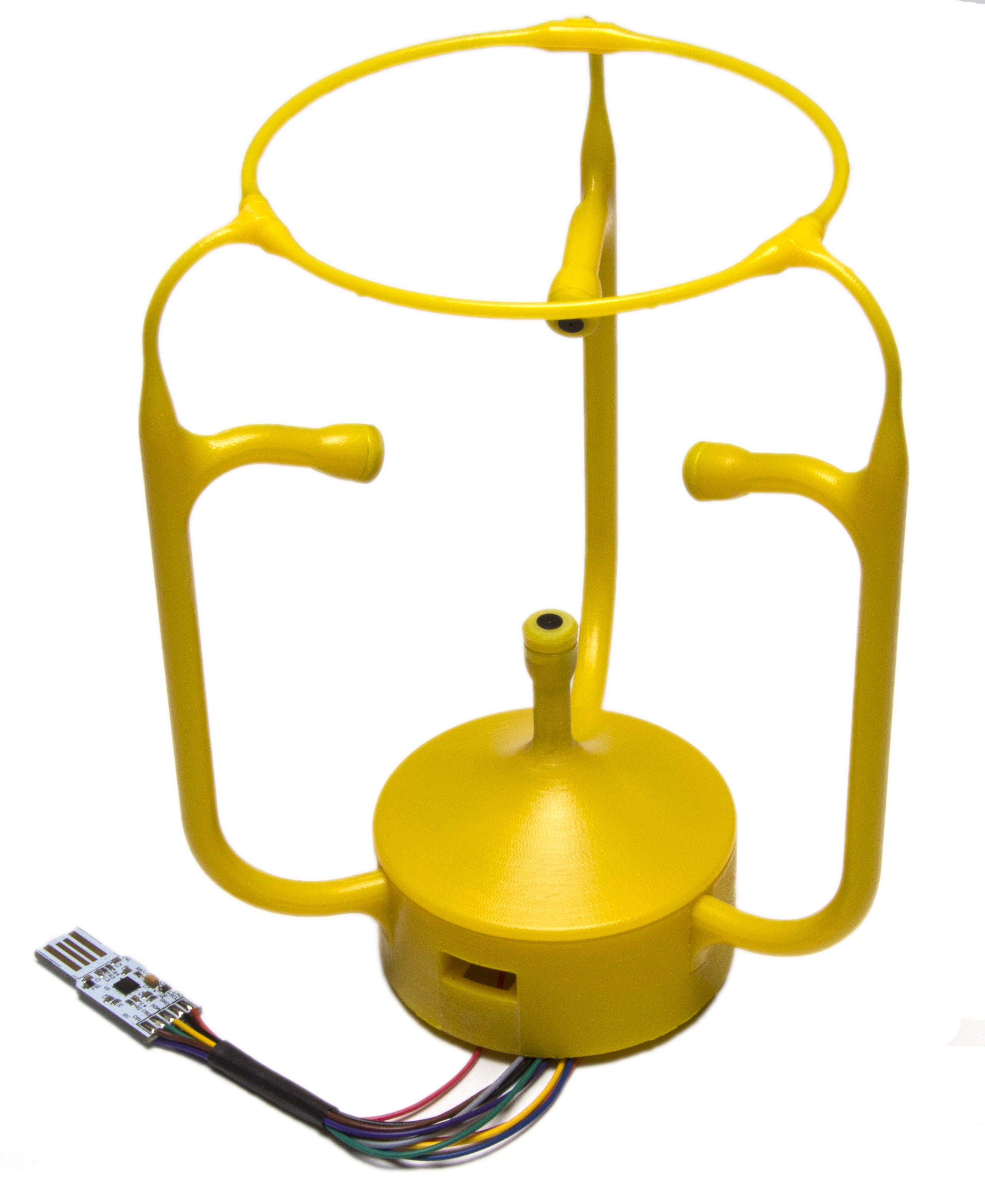}
        \caption{Anemometer}
    \end{subfigure}
    \begin{subfigure}[p]{0.70\linewidth}
        \includegraphics[width=0.48\linewidth]{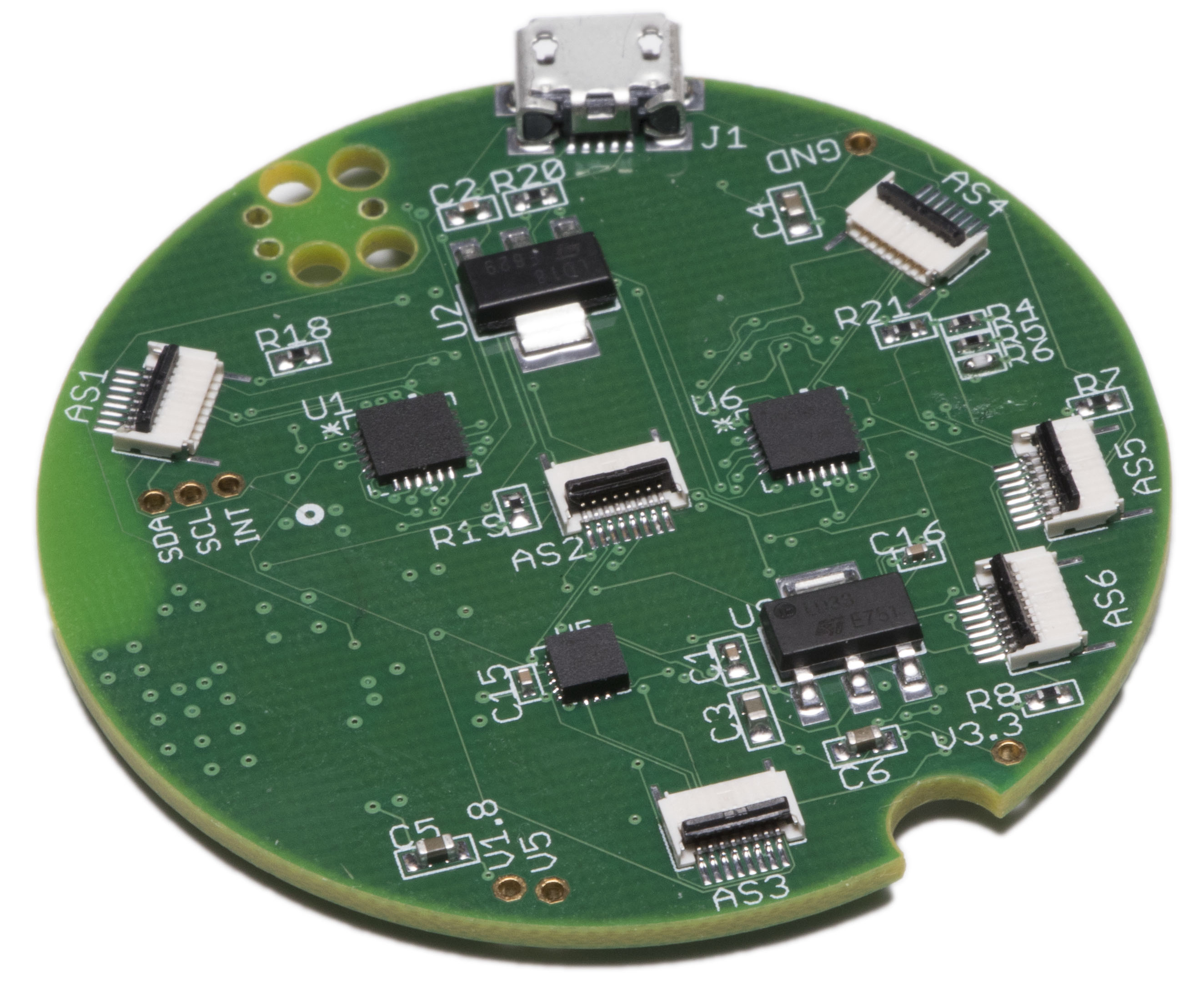}
        \includegraphics[width=0.48\linewidth]{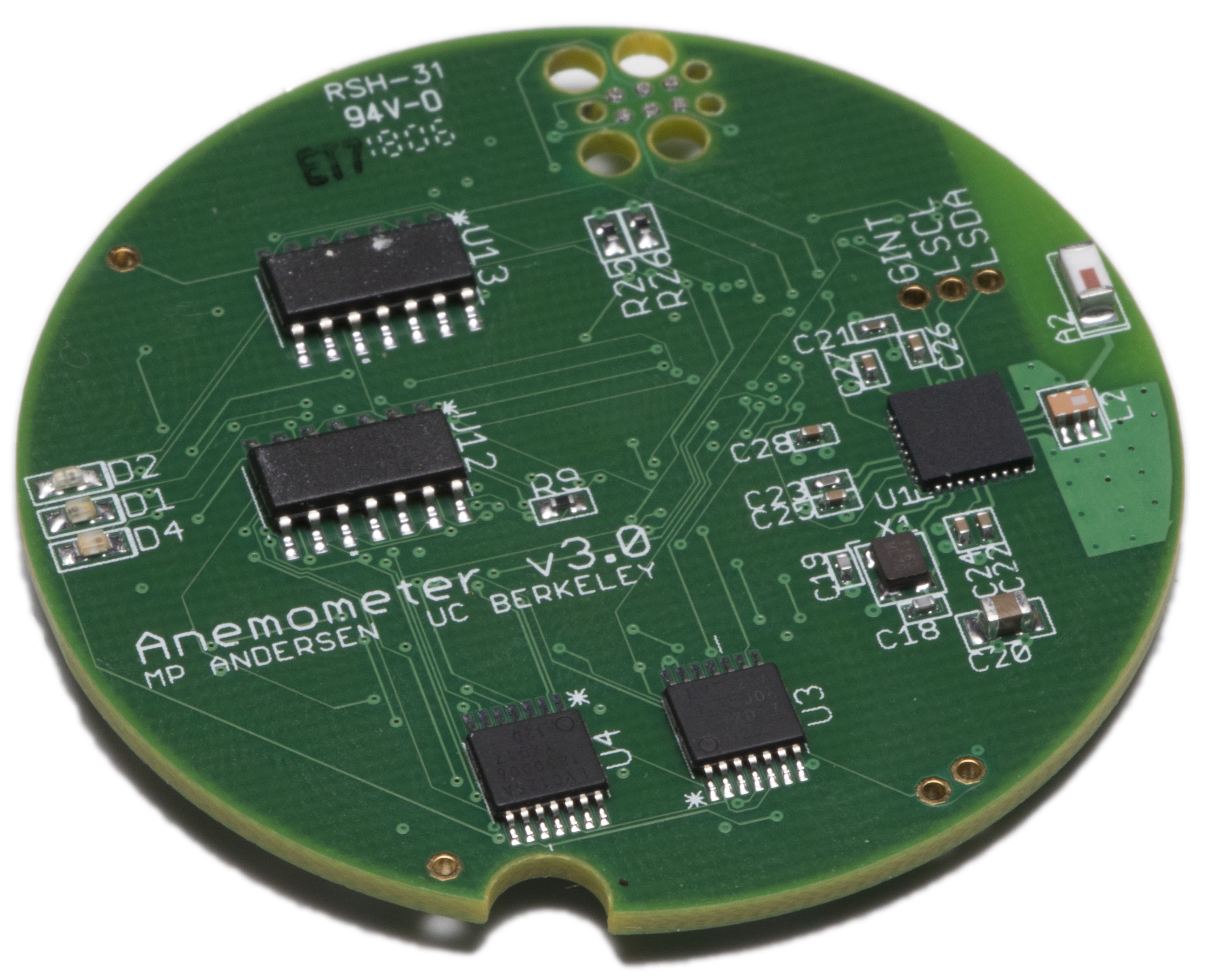}
        \caption{Hamilton-based PCB (bottom and top)}
    \end{subfigure}
    \caption{Hamilton-based ultrasonic anemometer}
    \label{fig:anemometer}
\end{figure}

We used anemometers based on the Hamilton platform~\cite{arens2020measuring}, each consisting of four ultrasonic transceivers arranged as vertices of a tetrahedron (Figure \ref{fig:anemometer}). To measure the air velocity, each transceiver, in turn, emits a burst of ultrasound, and the impulse is measured by the other three transceivers. This process results in a total of 12 measurements.

Calculating the air velocity from these measurements is computationally infeasible on the anemometer itself, because Hamilton does not have hardware floating point support and the computations require complex trigonometry. Measurements must be transmitted over the network to a server that processes the data. Furthermore, a specific property of the analytics is that it requires a contiguous stream of data to maintain calibration (a numerical integration is performed on the measurements). Thus, the application requires a high sample rate (1 Hz), and is sensitive to data loss. A protocol for \emph{reliable} delivery, like TCP or CoAP, is therefore necessary.

We note that the 1 Hz sample rate for this application is much higher than the sample rate of most sensors deployed in buildings. For example, a sensor measuring temperature, humidity, or occupancy in a building typically only generates a single reading every few tens of seconds or every few minutes. Furthermore, each individual reading from the anemometer is quite large (82 bytes), given that it encodes all 12 measurements (plus a small header). Given the higher data rate requirements of the anemometer application, we plan to use a higher-capacity battery than the standard AA batteries used in most motes. The higher cost of such a battery is justified by the higher cost of the anemometer transducers.

\end{document}